\documentclass[prb,aps,twocolumn,floats,superscriptaddress]{revtex4-2}
\include{header}
\usepackage{soul}
\usepackage{xcolor}
\usepackage{multirow}
\usepackage{manfnt}
\usepackage{amsmath}
\usepackage{graphicx}
\usepackage{amssymb}
\usepackage{epsfig}
\usepackage{simplewick}
\usepackage[colorlinks = true]{hyperref}
\begin{document}
	\newlength{\upit}\upit=0.1truein
\newcommand{\cube}{\mbox{\mancube}}
	\newcommand{\raiser}[1]{\raisebox{\upit}[0cm][0cm]{#1}}
	\newcommand{\ltappr}{{{\lower4pt\hbox{$<$} } \atop \widetilde{ \ \ \ }}}
	\newlength{\bxwidth}\bxwidth=1.5 truein
	\newcommand\frm[1]{\epsfig{file=#1,width=\bxwidth}}
	\newcommand{\cg}{{\cal G}}
	\newcommand{\dif}[2]{\frac{\delta #1}{\delta #2}}
	\newcommand{\ddif}[2]{\frac{\partial #1}{\partial #2}}
	\newcommand{\Dif}[2]{\frac{d #1}{d #2}}
	\newcommand{\str}{\hbox{Str}}
	\newcommand{\Str}{\underline{\hbox{Str}}}
	\newcommand{\tr}{{\hbox{Tr}}}
	\newcommand{\Tr}{\underline{\hbox{Tr}}}
	\newcommand{\dg}{^{\dagger }}
	\newcommand{\vk}{\mathbf k}
	\newcommand{\vq}{{\vec{q}}}
	\newcommand{\vp}{\bf{p}}
	\newcommand{\al}{\alpha}
	\newcommand{\BigL}{\biggl }
	\newcommand{\BigR}{\biggr }
	\newcommand{\gtappr}{{{\lower4pt\hbox{$>$} } \atop \widetilde{ \ \ \ }}}
	\newcommand{\si}{\sigma}
	\newcommand{\rarrow}{\rightarrow}
	\newcommand{\up}{\uparrow}
	\newcommand{\dw}{\downarrow}
	\def\fig#1#2{\includegraphics[height=#1]{#2}}
	\def\figx#1#2{\includegraphics[width=#1]{#2}}

	\newcommand{\bk}{{\bf{k}}}
	\newcommand{\bx}{{\bf{x}}}
\newcommand{\ba}{{\bf{a}}}
	\newcommand{\pmat}[1]{\begin{pmatrix}#1\end{pmatrix}}
	\newcommand{\ua}{\uparrow}
	\newcommand{\da}{\downarrow}
	
	\newcommand{\be}{\begin{equation}}
		\newcommand{\ben}{\begin{equation*}}
			\newcommand{\ee}{\end{equation}}
		\newcommand{\een}{\end{equation*}}
	\newcommand{\parr}{\parallel}
	\newcommand{\bmx}{\begin{array}}
		\newcommand{\emx}{\end{array}}
	\newcommand{\bean}{\begin{eqnarray*}}
		\newcommand{\eean}{\end{eqnarray*}}
	\newcommand{\dn}{^{\vphantom{\dagger}}}
	\newcommand{\nn}{\vphantom{-}}
	\newcommand{\lr}{\leftrightarrow}
	\newcommand{\ra}{\rightarrow}
	\newcommand{\la}{\leftarrow}
	\newcommand{\bb}[1]{\mathbb{#1}}
	\newcommand{\qqquad}{\qquad\qquad\qquad}
	\newcommand{\eps}{\epsilon}
	\newcommand{\sgn}[1]{{\rm sign}{#1}}
	\newcommand{\pref}[1]{(\ref{#1})}
	\newcommand{\tilda}[2]{\intopi{d{#1}}\Big(g_{#1}^A(\eps_p){#2}\Big)}
	\newcommand{\intpi}[1]{\int_{-\pi}^{+\pi}{#1}}
	\newcommand{\im}[1]{{\rm Im}\left[ #1 \right]}
	\newcommand{\trr}[1]{{\rm Tr}\Big[ #1 \Big]}
	
	\newcommand{\abs}[1]{\left\vert #1 \right\vert}
	\newcommand{\bra}[1]{\left\langle #1 \right\vert}
	\newcommand{\ket}[1]{\left\vert #1\right\rangle}
	\newcommand{\braket}[1]{\left\langle #1\right\rangle}
	\newcommand{\mat}[1]{\left(\bmx{cc}#1\emx\right)}
	\newcommand{\matc}[2]{\left(\bmx{#1}#2\emx\right)}
	\newcommand{\matn}[1]{\bmx{cc}#1\emx}
	\newcommand{\matl}[1]{\bmx{ll}#1\emx}
	\newcommand{\sepline}{\begin{center}\rule{8cm}{.5pt}\end{center}}
	\newcommand{\hi}{\noindent\currfilebase.pdf \hspace{.2cm}-\hspace{.2cm} \today}
	\newcommand{\hiy}{\noindent Yashar \currfilebase.pdf \hspace{.2cm}-\hspace{.2cm} \today}
	
	\setlength{\parindent}{0.5cm}
	\newcommand{\indentoff}{\setlength{\parindent}{0cm}}
	\newcommand{\EJK}[1]{{\color{olive} EJK: #1}}
	\newcommand{\imEJK}[1]{{\color{olive}\bf \Large \ddag}\marginpar{\scriptsize \color{olive}\bf  \ddag #1}}
	\newcommand{\SEC}[1]{{\color{blue} \textit{#1}}}
	\newcommand{\red}[1]{{\color{red}#1}}
	\newcommand{\blue}[1]{{\color{blue}#1}}
	\newcommand\relbd{\mathrel{{\bf\smash{{\phantom- \above1pt \phantom-
	}}}}}
	\newcommand\ltdash{\raise-0.7pt\hbox{$\scriptscriptstyle |$}}
	\def\fig#1#2{\includegraphics[height=#1]{#2}}
	\def\figx#1#2{\includegraphics[width=#1]{#2}}
	\newlength{\figwidth}
	\figwidth=10cm
	\newlength{\shift}
	\shift=-0.2cm
	\newcommand{\fg}[3]
	{
		\begin{figure}[ht]
			
			\vspace*{-0cm}
			\[
			\includegraphics[width=\figwidth]{#1}
			\]
			\vskip -0.2cm

			\caption{\label{#2}
				\small#3
			}
	\end{figure}}
	\newcommand{\fgb}[3]
	{
		\begin{figure}[b]
			\vskip 0.0cm
			\begin{equation}\label{}
				\includegraphics[width=\figwidth]{#1}
			\end{equation}
			\vskip -0.2cm
			\caption{\label{#2}
				\small#3
			}
	\end{figure}}
	\graphicspath{{Figures/CPTVfigs/}}
	
	\title{Breakdown of order-fractionalization in the CPT model
	}
	\author{Aaditya Panigrahi}
	\affiliation{
		Center for Materials Theory, Department of Physics and Astronomy,
		Rutgers University, 136 Frelinghuysen Rd., Piscataway, NJ 08854-8019, USA}
	\author{Alexei Tsvelik}
	\affiliation{Division of Condensed Matter Physics and Materials Science, Brookhaven National Laboratory, Upton, NY 11973-5000, USA}
	\author{Piers Coleman}
	\affiliation{
		Center for Materials Theory, Department of Physics and Astronomy,
		Rutgers University, 136 Frelinghuysen Rd., Piscataway, NJ 08854-8019, USA}
	\affiliation{Department of Physics, Royal Holloway, University
		of London, Egham, Surrey TW20 0EX, UK.}
	
	\date{\today}
	
	\pacs{PACS TODO}
	\begin{abstract}
		{ 
{We present an analysis of the half-filled CPT model, an analytically tractable Kondo lattice model with Yao-Lee spin-spin interactions on a 3D hyperoctagon lattice, proposed by Coleman, Panigrahi, and Tsvelik. Previous studies have established that the CPT model exhibits odd-frequency triplet superconductivity and order fractionalization. Through asymptotic analyses in the small $J$ and large $J$ Kondo coupling limits, we identify a quantum critical point at $J_c$, marking a transition from a superconductor to a Kondo insulator. By estimating the vison gap energy to account for thermal gauge fluctuations, we determine the energy scales governing the thermal breakdown of order fractionalization.   Moreover, at large 
$J$ the Kondo insulator undergoes orbital decoupling, leading to the formation of a decoupled Kitaev orbital liquid. These findings and analogies with the $\mathbb{Z}_2$-gauged $XY$ model lead us to propose a tentative phase diagram for the CPT model at half-filling.}
		}
	\end{abstract}
	\maketitle
	\section{Introduction}

			In heavy fermion materials, the coherent scattering of
		conduction electrons off a lattice of local moments
		produces a wide variety of emergent behavior. These range
		from heavy fermion metals and superconductors
		\cite{petrovic_heavy-fermion_2001,flint_heavy_2008,tsvelik_order_2022,coleman_solvable_2022,choi_topological_2018,seifert_fractionalized_2018}
		to topological Kondo Insulators
		\cite{dzero_topological_2010,neupane_surface_2013}. Broader
		classes of flat band systems, such as Moir\'e materials can
		also be modeled as heavy fermion systems\cite{song_magic-angle_2022,kumar_gate-tunable_2022,ramires_electrically_2018}.   
				
		\begin{figure}[h]
			
			\includegraphics[width=\columnwidth]{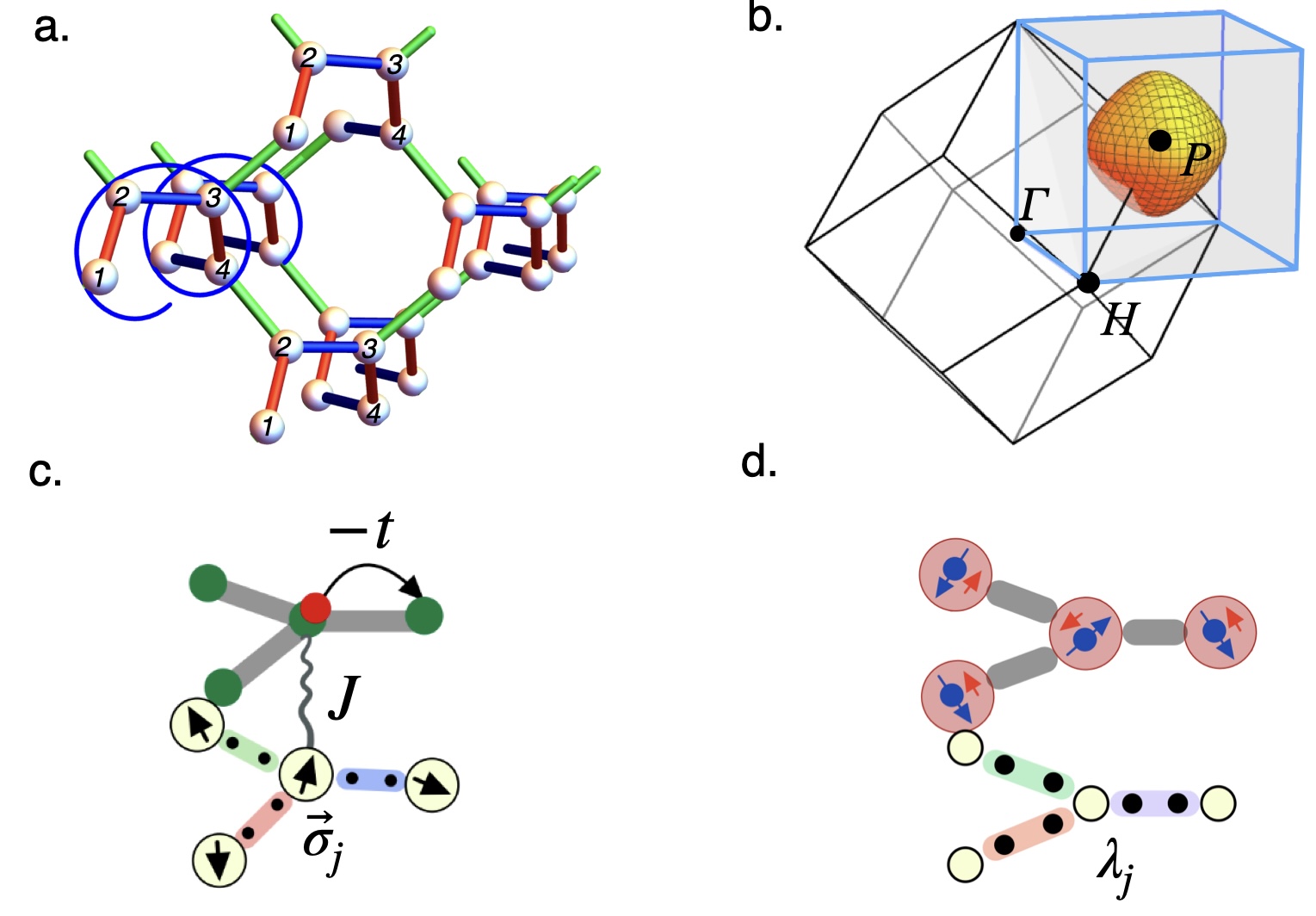} \caption{\small (a)
				The hyperoctagon lattice, a trivalent
				structure consisting
				of a spiral of four atoms per unit
				cell embedded in a body centered
				cubic lattice. 
				(b) A Yao-Lee spin liquid on this lattice results in 
				a flux-free gauge
				configuration, giving rise to a Majorana Fermi surface centered at the $P=(\pi,\pi,\pi)$ point in the Brillouin zone. 
				(c) Schematic showing the Kondo coupling between the conduction electrons
				and the local moments. (d)
				Cartoon illustrating the  ground-state
				of the large $J$ region of the CPT model,
				showing in pink, the formation of 
				Kondo singlets between conduction
				electrons and spins and 
				a decoupled Kitaev orbital liquid (KOL) .}  \label{KSLFig} 
			
		\end{figure}
		
			An insightful approach to heavy fermions is to consider them
		as the 
		Higgs phase\cite{coleman_1_1983,read_solution_1983,
			kogut_introduction_1979,vojta,Coleman:849264} of an underlying
		spin-liquid.  From this perspective, alternate patterns of spin
		fractionalization may drive new kinds of physics, such as pairing
		beyond the BCS paradigm.  One way to explore this idea is to study
		Kondo lattices with a pre-formed spin liquid. Attempts
		have focused on understanding Kondo lattice models, where the
		underlying spins interact via Kitaev
		interactions\cite{kitaev_anyons_2006,seifert_fractionalized_2018,choi_topological_2018}. Unfortunately,
		Kitaev-Kondo models lose their exact solvability as a
		result of the Kondo term: to avoid this difficulty, Coleman,
		Panigrahi, and Tsvelik (CPT) have recently proposed a three-dimensional Kondo
		lattice model,
		\cite{coleman_solvable_2022,tsvelik_order_2022}(Fig. \ref{KSLFig}) in
		which a Yao-Lee orbital-spin interaction\cite{yao_fermionic_2011}
		restores the solvability of the Kondo lattice at half filling.

		The Hamiltonian for the CPT model on a hyperoctagon lattice
		(Fig \ref{KSLFig}),
		\begin{equation} H_{CPT}=H_c+H_{YL}+H_K
			\label{CPTHam} \end{equation}
		 has three components:
		  \begin{align}
			H_{c}=&-t\sum_{\braket{i,j}} ( c\dg_{i\si}c_{j\si}+{\rm H.c.})-\mu\sum_{i}c\dg_{i\si}c_{i\si},
			\\H_{YL}=&K/2\sum_{\braket{i,j}}\lambda^{\al_{ij}}_i\lambda^{\al_{ij}}_j
			(\vec{S}_i\cdot\vec{S}_j),\label{YLH}
			\\H_{K}=&J\sum_{i}(c\dg_{i}\vec{\si}c_{i})\cdot \vec{S}_{i}
			\label{CPTComp} \end{align} 
		Here $\langle i,j\rangle $ are neighboring sites on the
		hyper-octagonal lattice\cite{Hermanns}, a  trivalent body centered cubic (BCC) crystal 
		with four  atoms per primitive unit cell, coiled around a helix
		to form alternating square and octagonal
		spirals(Fig. \ref{KSLFig}b).
		Each site supports both electrons
		$c_{i\si}$, localized spins $\vec{S}_{i}$ and localized orbital
		$\vec{\lambda}_i$ degrees of freedom. The conduction term $H_c$ describes
		electrons hopping between nearest neighboring sites. The Kondo interaction
		$H_K$ antiferromagnetically couples the conduction electrons to spins
		$\vec{S}_i$ at each site. Finally, the Yao-Lee term couples the orbitals
		$\vec{\lambda}_i$ via a Kitaev-like anisotropic interaction,
		``decorated''
		by a Heisenberg coupling between the nearest neighbor spins $\vec{S}_i$. The
		$\alpha_{ij}$ are the Ising coupling of orbital components along the
		$\alpha_{ij}=x,y,z$ bond
		directions(Fig. \ref{KSLFig}c).

		The anisotropic Ising coupling between orbitals induces
		Majorana fractionalization \cite{yao_fermionic_2011} of spins $\vec{S}_{j}=- (\frac{i}{2})\vec{\chi}_j\times\vec{\chi}_j$ and orbitals $\vec{\lambda}_j=i \vec{b}_j\times \vec{b}_j$. In the physical Hilbert space, where $\si^a_j\lambda^{\al}_j=2i \chi^a_j b^{\alpha}_j$, the fractionalized form of the Yao-Lee Hamiltonian
		$H_{YL}$ \eqref{YLH} is
		\begin{equation}
			H_{YL}=K\sum_{\braket{i,j}}\hat{u}_{ij}(i\vec{\chi}_i\cdot \vec{\chi}_{j}).
		\end{equation}
		Here, $\hat{u}_{ij}=i b^{\al_{ij}}_ib^{\al_{ij}}_j$ are the  static
		$\mathbb{Z}_2$ gauge fields, (i.e. $[H_{YL},\hat{u}_{ij}]=0$).

		In three dimensions, $\mathbb{Z}_{2}$ gauge theories undergo a finite
		temperature Ising phase transition at $T_{c1}$, into a deconfined phase, in
		which the visons (plaquettes with a $\pi$ flux) are
		linearly confined.  In the Yao Lee model on a hyper-octagonal lattice, 
		$T_{c1}\sim
		0.036K$\cite{hermanns_physics_2018,obrien_classification_2016,coleman_solvable_2022},
		leading to a fractionalization of the spins into majoranas at lower
		temperatures $T<T_{c1}$. 
		
			\begin{figure}[h]
\includegraphics[width=\columnwidth]{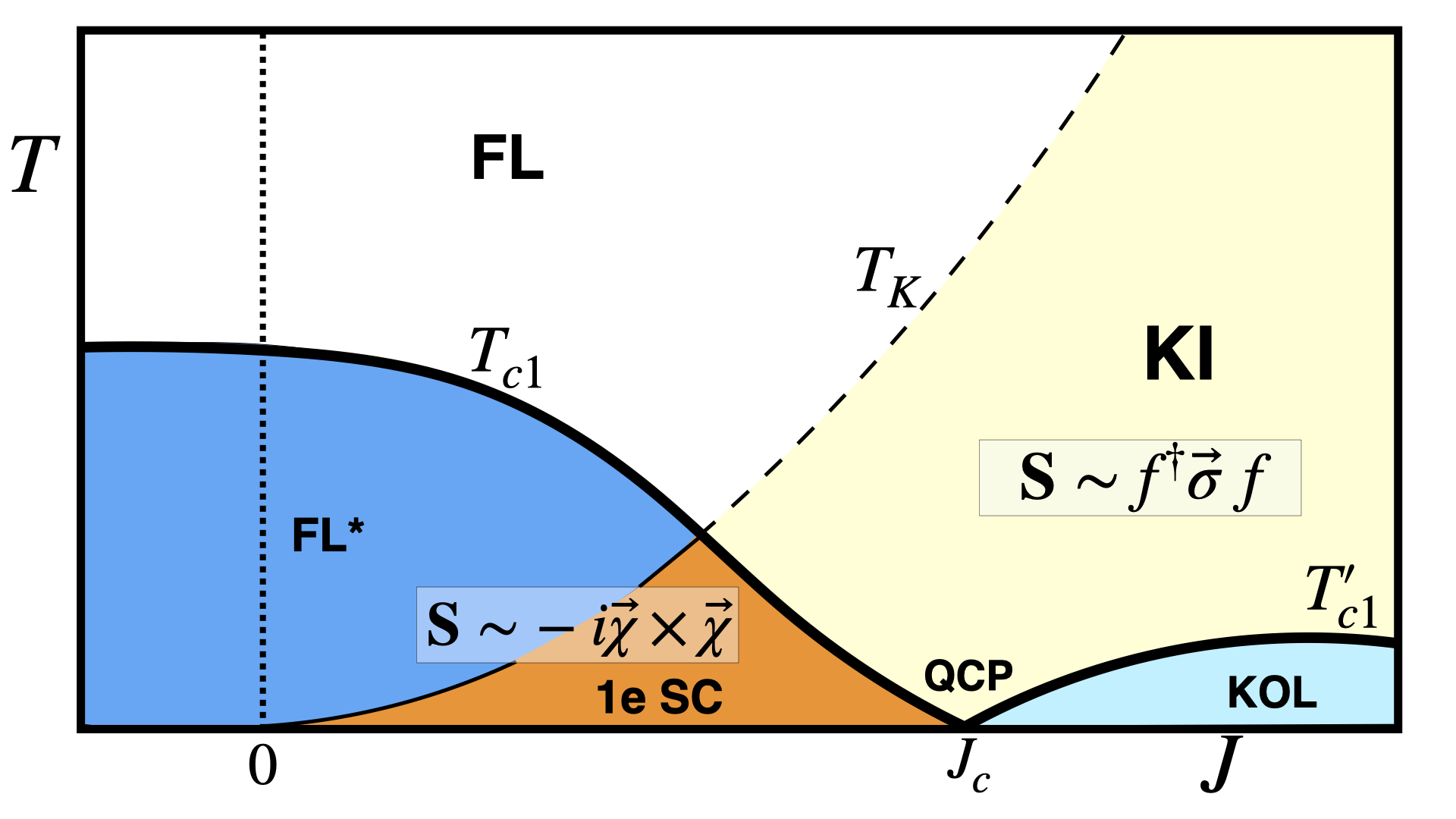}
\caption{\small Proposed phase diagram of the half-filled CPT model
half-filling. A finite temperature Fermi liquid (FL) with a small Fermi
surface(FL) develops at small $J$, with a cross-over into a Kondo
insulating phase{KI} at large $J$.  For $T<T_{c1}$ and small $J$, the
spins fractionalize into Majorana ($\vec{\chi}$) fermions, forming a
3D Yao Lee spin liquid.  A logarithmic divergence in the pairing
susceptibility at small positive $J$ gives rise to a charge $1e$,
$S=1/2$ electron-majorana condensate (1eSC), with a transition temperature
comparable to the single-ion Kondo temperature $T_{K}$.
				Once $T_{K}$ exceeds $T_{c1}$, $T_{c1}$ 
				drops to zero at $J=J_{c}$, giving rise to a
				superconductor-insulator quantum
				critical point (QCP),  forming 
				a Kondo insulator (KI) where the spins fractionalize as
				Dirac ($f$) fermions.  For $T<T'_{c1}$, the Kondo
				insulator co-exists with a decoupled Kitaev orbital-liquid (KOL). 
			}
			\label{PhaseDiagram}
		\end{figure}

		In this paper, we examine the phase diagram
		(Fig. \ref{PhaseDiagram}) of the CPT model at
		half-filling. Central to the
		phase diagram are the two characteristic energy scales: the Kondo temperature $T_K$ and the
		Ising transition temperature $T_{c1}$. Below the Kondo
		temperature $(T<T_K)$ the conduction electrons screen
		the spins, while below the Ising temperature
		$(T<T_{c1})$ the spins fractionalize
		$\vec{S}_j=-\frac{i}{2}\vec{\chi}_j\times
		\vec{\chi}_j$ into $\vec{\chi}_j$ majoranas. The
		interplay between spin fractionalization and Kondo
		screening determines the phase diagram. 
		
		Our proposed phase diagram  for the half-filled CPT
		model (CPTM), shown in Fig. \ref{PhaseDiagram}, 
		is based on the analytical tractability of the extreme
		limits of small and large $J$ Kondo coupling, where
		the model model exhibits superconductivity and Kondo
		insulating behavior, respectively. 
One or more 
quantum phase transitions at an intermediate coupling $J_{c}$ must
therefore separate these distinct phases.
		
		Applying Ockham’s razor, we propose a single quantum
critical point (QCP) defining the superconductor-insulator
transition. While we cannot entirely rule out an alternative
first-order quantum phase transition between the two phases, this
scenario would necessitate a finite temperature critical endpoint,
with a thermal rather than a quantum phase transition governing the
changing pattern of spin fractionalization.

		The Kondo temperature $T_K(J)$ and the Ising
		transition temperatures $T_{c1}(J)$ and $T_{c1}' (J)$
		divide the phase diagram (Fig. \ref{PhaseDiagram})
		into five phases, which we summarize below:
		\begin{enumerate}
			\item {\sl Fermi Liquid} (FL) ($5t \gg
			T>T_{c1}$ and $T>T_K$): At temperatures much
			higher than $T_{K}$ and $T_{c1}$, but much
			lower than the conduction band-width $W\sim
			5t$ the CPT model forms a conduction Fermi
			liquid, weakly coupled to its embedded spins
			and orbitals. The spins and orbitals will exhibit paramagnetic behavior characterized by Curie-Weiss susceptibility.  
			\item {\sl Fermi Liquid*} (FL$^{*}$) ($T<T_{c1}$ and $T>T_K$): Below the Ising transition temperature $(T<T_{c1})$ the spins fractionalize into Majorana fermions $\vec{S}_j=-\frac{i}{2}\vec{\chi}_j \times \vec{\chi}_j$, forming a Yao-Lee spin liquid with a Majorana Fermi surface. Since the system is above the Kondo temperature ($T>T_K$), the Yao-Lee spin liquid remains unscreened by the conduction sea. Thus, in this phase, the CPT model consists of a decoupled Fermi liquid and a Yao Lee spin liquid, constituting a Fermi liquid* phase.
			\item {\sl 1e Superconductor} ($1e$ SC) ($T<T_{c1}$ and $T<T_K$): In this phase, spins undergo Majorana fractionalization and experience Kondo screening by conduction electrons. The precise nesting between the Majorana Fermi surface of the Yao-Lee spin liquid and the conduction sea at half-filling results in a Peierls-like logarithmic instability between the electron and Majorana Fermi surface for an infinitesimal Kondo coupling \cite{coleman_solvable_2022}.
			
			This results in an odd-frequency triplet superconductor characterized by a fractionalized charge $e$-spinor order in the small $J$-limit \cite{komijani_order_2018,tsvelik_order_2022,coleman_solvable_2022}. The system exhibits a neutral Fermi surface due to an imbalance between $4$ conduction Majoranas and $3$ spin liquid Majoranas. For more details see {section \ref{sec2}}. 
			
			\item {\sl Kondo Insulator} (KI) ($T_{K}>T>T_{c1},T'_{c1}$
			Below the Kondo temperature
			$T_K$, but at temperatures larger than the
			Ising temperatures $T_{c1}, T'_{c1}$, 
the spins are screened by the
			conduction sea, forming a Kondo insulator,
			while the decoupled orbitals are
			unfractionalized. 

\item {\sl Kitaev Orbital Liquid} (KOL) 
The ground state properties of the Kondo insulator at large $J$ can be
determined by carrying out a strong coupling expansion in $1/J$.
In this limit, the ground-state is 
			a product ground state of
			local Kondo singlets ($(\uparrow\Downarrow - \downarrow
			\Uparrow)_{j}$), 
			co-existing with a  degenerate manifold of decoupled 
			orbitals $\lambda^{\alpha }_{j}$
			\begin{equation}\label{largeJ}
				\vert \Psi \rangle  = \prod_{j} (\uparrow\Downarrow - \downarrow \Uparrow)_{j}\vert \{\lambda_{j}\}\rangle.
			\end{equation}
			The large spin-gap of order $J$ in this phase allows a treatment of the
			Heisenberg bond operator $\vec{S}_{i}\cdot \vec{S}_{j}\sim
			\langle \vec{S}_{i}\cdot \vec{S}_{j}\rangle  \sim O (t/J)$.
			as a static variable. Through the Yao-Lee coupling, this
			lifts the 
			orbital degeneracy, giving rise to a decoupled 3D Kitaev
			orbital liquid (orbital analog of Kitaev spin liquid) below the 
			second Ising transition temperature $T'_{c1}$. For more
			details see {section \ref{sec3} }. 

		\end{enumerate}

		This paper is structured as follows:  in
		section \ref{sec2} we summarize
the small-$J$ limit of the CPT model; in section \ref{sec3} we 
		present the large-$J$ limit, while in the final
section \ref{sec4} we discuss the intermediate coupling 
behavior of the CPT model.


	\section{Small-$J$ limit of the CPT model}\label{sec2}
	For completeness, here we summarize the results of our previous
	work\cite{coleman_solvable_2022}  on the half-filled CPT
	model in the analytically tractable small $J$ limit.
	  
	We recall that the CPT model is a Kondo lattice model with a
quartet state at each site, comprised of a $S=1/2$ spin and orbital
degree of freedom at each site. The Yao-Lee interaction between sites
acts to establish an emergent static $\mathbb{Z}_{2}$
field\cite{tsvelik_order_2022,coleman_solvable_2022}.   At half
filling, the CPT model develops nested
electron and Majorana Fermi surfaces and for infinitesimal Kondo
coupling, the system undergoes a second-order phase transition into a
spinor-ordered electron-majorana
condensate\cite{coleman_solvable_2022}. 
	
	{\it Yao Lee spin-liquid}: The three dimensional Yao Lee (3DYL) model on a hyperoctagon lattice shares many of the properties of a 2D Kitaev
spin liquid, most notably, the presence of gapped $Z_{2}$ flux
excitations, 
described by Wilson loops - products of
the gauge fields $W= \prod u_{(i,j)}= \pm 1  $ around closed
ten-fold loops of the hyperoctagon lattice(where $(i,j)$ orders the sites $i$ and $j$ 
along  $xx,\ yy$ and $zz$ bonds so that
the site furthest in the $y$, $z$ and $x$
directions respectively, is placed first \cite{Hermanns}). In the spin liquid ground-state,
all loops are trivial $W=1$\cite{Hermanns}; flipping the sign of a
Wilson loop creates a flux excitation (vison), with an energy
determined as a fraction of $K$.  Unlike 2D Kitaev spin liquids,
the 3DYL undergoes an Ising phase transition at $T_{c1}\sim 0.036K$
into a Higgs phase where the elementary $\mathbb{Z}_2$ gauge excitations (visons) are 
linearly confined 
\cite{janssen20,mishenko17,eschmann_thermodynamic_2020,coleman_solvable_2022}. 
and
the Majorana fields describe coherent, fractionalized
spin excitations.

Below the Ising transition temperature $T_{c1}$, the confinement
of visons in the 3D Yao Lee model  allows a gauge choice $u_{i,j}=1$
\cite{hermanns_physics_2018} 
leading to a translationally invariant Hamiltonian. 
Transforming to
a momentum basis, 
	\begin{equation} \vec{\chi}_{{\bf
k},\alpha }=\frac{1}{\sqrt{N}}\sum_{j}\vec{\chi}_{j\alpha} e^{-i{\bf
k}\cdot {\bf R}_j},
\label{MajoranaFT} \end{equation} 
where ${\bf R}_{j}$ is the position of the unit-cell in the BCC
lattice,   
$N$ is the number of primitive unit cells in the lattice and $\alpha \in [1,4]$
is the site index within each unit cell. 
	The ground state Hamiltonian for the Yao-Lee spin liquid 
is then
	\begin{equation}
		H_{YL}=K 
		\sum_{{\bf k}\in \cube }
		{\vec{\chi}}^{\dagger}_{\bf{k},\al}
		h({\bf k})_{\alpha,\beta}{\vec{\chi}}_{\bf{k}\beta},
		\label{YLHam}
	\end{equation}
where $\alpha,\beta\in [1,4]$ are the site indices and
		\begin{equation}
		\small
		h(\bf{k})=\begin{pmatrix}
			0 && i && i e^{-i \bf{k}\cdot \bf{a}_2} && i e^{-i \bf{k}\cdot \bf{a}_1} 
			\\-i && 0 && -i && i e^{-i \bf{k}\cdot \bf{a}_3}
			\\-i e^{i \bf{k}\cdot \bf{a}_3} && i && 0 && -i
			\\-i e^{i \bf{k}\cdot \bf{a}_1} && -i e^{i \bf{k}\cdot \bf{a}_2} && i && 0
		\end{pmatrix}
	\end{equation}
where 
$	\ba_1=(1,0,0);\ba_2=\frac{1}{2}(1,1,-1);\ba_3=\frac{1}{2}(1,1,1),$
are the primitive BCC lattice vectors.
Since $\chi_{-\bk} = \chi\dg_{\bk}$, the momentum sum is restricted to
half
the Brillouin zone, corresponding to a cube 
($\cube$) of side length $2\pi$ centered at the
$P$ point at  $(\pi,\pi,\pi)$.
The spectrum $E_{\bk }\equiv K \epsilon (\bk )$,
determined by ${\rm det} [\epsilon \underline{1}-h ({\bk })]=0$, or
\begin{equation}\label{eqn: characteristic}
\epsilon^4-6\epsilon^2 -8\epsilon (s_{x}s_{y}s_{z})
+ [9- 4 (s_{x}^{2}+s_{y}^{2} + s_{z}^{2})]=0,
\end{equation}
(where $s_{l}\equiv \sin (k_{l}/2)$, $l=x,y,z$),
contains a {\sl single } Fermi surface where $\epsilon=0$ and
$s_{x}^{2}+s_{y}^{2} + s_{z}^{2}=9/4$, centered at $P$ \cite{Hermanns2}.
(Fig. \ref{KSLFig} b.)
	
	{\it Conduction electrons}: The nesting between Majorana and
	conduction Fermi surface becomes evident upon performing the
 gauge transformation  $(c_1,c_2,c_3,c_4)_{\vec{R}}\rightarrow e^{i{\bf (\pi,\pi,\pi)\cdot R}} (c_1,i c_2,c_3,- i c_4)_{\vec{R}} $ on conduction electrons in the unit-cell at $\vec{R}$. The resulting conduction Hamiltonian $H_c$ takes the form,
	\begin{equation}
		H_c=\sum_{{\bf k}\in {\bf BZ}}c^{\dagger}_{{\bf k},\si\al}[-t\  h({\bf k})-\mu \mathbb{I}]_{\al\beta}c_{{\bf k},\si\beta},
		\label{CondHam}
	\end{equation}
	where $\alpha,\beta $ denote the site indices of the unit cell, 
	thus sharing the same hopping 
	matrix $h (\bk )$ 
	(\ref{YLHam}) as the spinons in the 3D Yao Lee spin liquid. At
	$\mu=0$, the conduction sea develops an electron and a hole Fermi
	surface at $P$ and $-P$ respectively, which can be rewritten as four
	Majorana Fermi surfaces centered at $P$. These Fermi surfaces perfectly nest with the
 three Majorana Fermi surface of the spin liquid, which facilitates a BCS-like mean-field treatment of
	the Kondo interaction.
	
	{\it Kondo interaction}: To this end, we express the Kondo interaction in Majorana spin representation as,
	\begin{equation}
		H_{K}=
J\sum_{j}(c\dg_{j}\vec{\sigma}c_j)\cdot (-\frac{i}{2}
\vec{\chi}_j\times \vec{\chi}_j)\equiv - \frac{J}{2}\sum_{l}c\dg_{j}
(\vec{\chi }\cdot \vec{ \sigma })^{2}c_{j}
	\end{equation}
	
	{\it Small-$J$ limit}: Below the Ising transition $T_{c1}$ a Hubbard-Stratonovich transformation of the Kondo interaction in terms of the charge-$e$ spinor order  $V_j=-J\langle \vec{\chi}_a\cdot \vec{\si} c_j\rangle=(V_{j\up},V_{j\dw})^{T}$,  
	\begin{equation}
		H_{K}=\sum_{j} \bigl[c\dg_{j} (\vec{\si}\cdot \vec{\chi}_j)V_j+{\rm H.c}\bigr] +\frac{2 V\dg_j V_j}{J}.
		\label{MeanFieldHamiltonian}
	\end{equation}
	The uniform saddle point of the Hamiltonian then provides us 
	with the mean-field
	solution\cite{tsvelik_order_2022,coleman_solvable_2022} in the
	small $J$ limit. The mismatch in the number of conduction (4) and spin liquid majoranas (3) in equation (\ref{MeanFieldHamiltonian}) results in one conduction majorana remaining gapless as the fractionalized order condenses.

	In the mean-field solution with uniform spinor configuration $V_{j\sigma}=(\text{V}/\sqrt{2})z_{\sigma}$, the electronic self-energy in momentum space is expressed as:
	\begin{equation}
		\Sigma_{\mathbf{k},\omega}=(1-\mathcal{Z}\otimes \mathcal{Z}^{\dg})V^{2}D(\mathbf{k},\omega)
	\end{equation}
where $\mathcal{Z}=[z,i \si_y z^{*}]^{T}$ is the spinor order in
Balian-Werthamer notation, and $D(\mathbf{k},\omega)=[\omega-K
h(\mathbf{k})]^{-1}$ is the spin-liquid propagator. The matrix 
$\mathcal{Z}\otimes \mathcal{Z}^{\dg}$ projects out one Majorana component of
the conduction sea, leaving three components of the conduction
sea to hybridize with the spin-liquid, gapping them out in a fashion similar
to a Kondo insulator. The component of the conduction sea projected
onto the spinor ${\cal Z}$ forms a gapless ``neutral'' Fermi surface. 

The superconducting nature of the spinor order phase is evident when
expressing the self-energy in terms of three orthogonal d-vectors,
formed from bilinears of the spinors, \cite{coleman_solvable_2022}:
\begin{equation}
	\hat{\bf{d}}^1+i \hat{\bf{d}}^2=z^{T}(-i \si_2)\mathbf{\si} z, \hat{\bf{d}}^3=z^{\dg}{\bf{\si}}z.
\end{equation} 
Written in terms of the d-vectors, the self-energy separates 
into magnetic and pairing components:
\begin{equation}
	\Sigma=\Sigma_N({\bf k},\omega)+\Delta ({\bf k},\omega) \tau_+ +\Delta^{\dg} ({\bf k},\omega) \tau_-,
\end{equation}
which describe 
the coexistence of odd-frequency magnetism and triplet superconducting
order, where
$$\Sigma_{N}({\bf k},\omega)=\frac{1}{4}(3-(\hat{\bf d}^3\cdot {\bf
\sigma})\tau_3)\Sigma_0({\bf k},\omega)$$ describes odd-frequency
magnetism and 
$$\Delta({\bf k},\omega)=-\frac{1}{4}\bigl [(\hat{\bf d}^1+i\hat{\bf
d}^2)\cdot {\bf \sigma}\bigr ]\Sigma_0({\bf k},\omega), $$ describes
the triplet superconductivity.  Here 
\begin{equation}\label{}
\Sigma_{0} ({\bf k} ,\omega) = V^{2}D ({\bf k},\omega)
\end{equation} 
describes the hybridization with the spin liquid. On the Fermi surface
of the spin liquid, $\Sigma_{0} ({\bf k}_{F},\omega )\sim
\frac{1}{\omega}$  is an odd function of frequency. 
The complex d-vector $\hat{\bf{d}}^1+i \hat{\bf{d}}^2$ breaks the time-reversal symmetry, representing the two component superconducting order in the small-$J$ limit.

	The exactness of the small-$J$ limit is a consequence of a
logarithmic divergence in the pairing susceptibility, which results
in a second-order phase transition at a transition temperature (Fig. \ref{PhaseDiagram}),
	\begin{equation}
		T_{K}= W\exp\left(-\frac{1+K/t}{\rho J}\right).
		\label{TransitionTemperature}
	\end{equation}
	At temperatures $T<T_{K}$,
	the system forms an 
	electron-majorana pair condensate with a spinor order parameter. 
	Here, small $J$, combined with the logarithmic divergence of the
	spinor pair susceptibility, 
	acts as the small parameter for the mean-field
	theory. Further, 
this spinor order generates odd-frequency triplet pairing within the
conduction sea. 
	
	
	\section{Large $J$ limit of the CPT Model}\label{sec3} The large $J$ limit
of the half-filled CPT model can be solved in a strong coupling
expansion in $1/J$. In this limit the half-filled ground state is a product state of local
singlets of electrons and spins at each site
(Fig. \ref{KSLFig}d), forming a Kondo insulator 
i.e., \begin{equation}\label{largeJ} \vert \Psi \rangle = \prod_{j}
(\uparrow\Downarrow - \downarrow \Uparrow)_{j}\vert
\{\lambda_{j}\}\rangle.  \end{equation} 
where the ket $\vert
\{\lambda_{j}\}\rangle$ describes an arbitrary configuration of the orbital degrees of
freedom. There is a 
gap $\Delta_{g}=J$ between the singlet ground-state and the 
triplet excited states which allows
us to carry out
	a strong coupling expansion in powers of $K/J$ and $t/J$.
	
	\begin{figure}[h]
		\includegraphics[scale=.125]{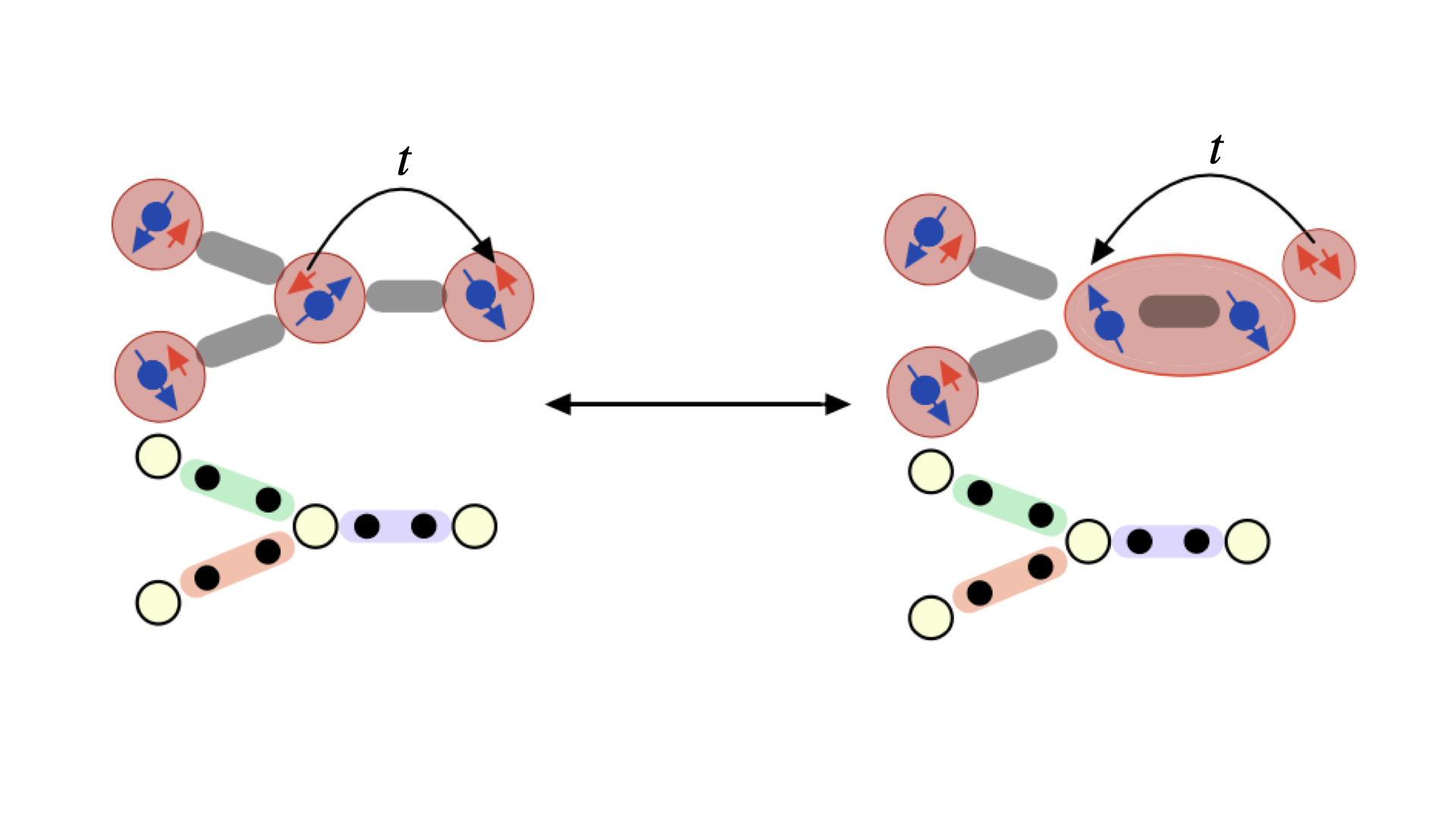}
		\caption{Depicts the Kondo insulator phase of the CPT model. The ground state is a product state of local electron-local spin singlets, resulting in decoupled orbitals. Through the virtual hopping of conduction electrons between nearest neighbors, the Kondo singlet breaks into nearest neighbor holon-doublon virtual excitation, and the spins form a singlet. The system reverts to its original state via back-hopping to form a product state of Kondo singlets. This virtual excitation leads to the orbitals fractionalizing into Majorana, resulting in a decoupled Kitaev orbital liquid. }
		\label{LargeJ}
	\end{figure} 
	
	To first order in the  expansion, 
	$\langle \vec{S}_i \cdot \vec{S}_j\rangle=0$ and the
	expectation value of the Yao Lee term \eqref{YLH} is zero.  However, 
electron hopping 	(Fig. \ref{LargeJ}) 		
gives rise to first-order corrections to the ground state:
	\begin{equation}
		\ket{\Psi}=\ket{\Psi_0}+\sum_{j}\frac{\ket{j}\bra{j}\hat{H}_{c}\ket{\Psi_0}}{E_0-E_j},
	\end{equation}
	where, 
	\begin{equation}
		\hat{H}_{c}=-it\sum_{\braket{i,j}}(c^{\dagger}_{i\si}c_{j\si}- H.c).
	\end{equation}
Written out explicitly, 
	\begin{equation}
		\small	\ket{\Psi\{\lambda_j\}}=\left(1-\frac{2}{3J}\left[-it\sum_{\braket{i,j}}(c^{\dagger}_{i\si}c_{j\si}- H.c)\right]\right)\ket{\Psi_0\{\lambda_j\}}.
	\end{equation}
The hopping moves one electron to a neighboring site forming a doubly
occupied singlet state, leading to two unscreened local moments $\vec{
S}_{i}$ and $\vec{S}_{j}$.  Since the hopping 
preserves the spin-singlet character of the wavefunction, the two
unscreened neighboring spins must form a singlet with $\vec{S}_{i}\cdot\vec{S}_j=-\frac{3}{4}$.
Thus to leading order 
	\begin{equation}
		\bra{\Psi\{\lambda_j\}}\vec{S}_{i}\cdot\vec{S}_j\ket{\Psi\{\lambda_j\}}=2\times
\left(-\frac{3}{4} \right)
\times\left(\frac{2t}{3J}\right)^{2}=-\frac{2}{3}\frac{t^2}{J^2}.
	\end{equation}
In the low energy degenerate manifold of orbital states, 
the matrix elements of the Yao-Lee interaction are 
	\begin{align}
		\nonumber\bra{\Psi\{\lambda_j\}}H_{YL}\ket{\Psi\{\lambda_j\}}&=\frac{K}{2}\sum_{\braket{i,j}}\braket{\vec{\si}_i\cdot\vec{\si}_j}\lambda^{\alpha_{ij}}_i\lambda^{\alpha_{ij}}_j
		\\ &=\frac{K^*}{2}\sum_{\braket{i,j}}\lambda^{\alpha_{ij}}_i\lambda^{\alpha_{ij}}_j,
	\end{align}
	with a renormalized coupling 
	\begin{equation}
		K^*=-\frac{8}{3}\frac{t^2}{J^2}K.
	\end{equation}
It follows that in the large Kondo coupling $J$ limit, the ground state of the CPT model at half-filling is a product state of spin-singlets with a decoupled Kitaev orbital liquid.  
	
	\section{Discussion}\label{sec4}

We now discuss the nature of the phase diagram at intermediate
coupling.  
At small $J$, 
the localized spins fractionalizes into vector ($S=1$) majoranas, 
$\vec{S}_j=-\frac{i}{2}\vec{\chi}_j\times\vec{\chi}_j$, while at large
$J$, they fractionalize as Dirac fermions
$\vec{S}=f^{\dg}_j\vec{\si}f_j$.
As the ground-state evolves from a small $J$
superconductor to a large $J$ Kondo insulator, 
we anticipate a superconductor-insulator transition. 
For reasons discussed below,
we expect this transition to be continuous 
providing a new example of a deconfined quantum
critical point, involving a transition from a 
$\mathbb{Z}_2$ Higgs $1e$
superconductor to a $\mathbb{Z}_2$ deconfined Kitaev orbital liquid.

There are a number of motivating arguments for a quantum-critical 
superconducting-insulator transition. 
Firstly, the transition involves a Fermi-surface transformation,
from  a conduction neutral Majorana Fermi surface in the superconductor 
to a Majorana Fermi surface of orbital excitations: this
is reminiscent of the continuous transitions from a small to a
large Fermi surface thought to occur in heavy fermion
criticality\cite{coleman_how_2001,senthil_weak_2004,paschen_hall-effect_2004}.
Secondly, energetic arguments tell us that a 
transition from superconductor to orbital Kondo insulator 
will occur when 
magnetic energy of the spin liquid ($E_{mag} \sim
-K$) drops below the Kondo energy gain ($E_{Kondo} \sim \frac{V^2}{W}$),
resulting from the partial gapping of the conduction Fermi surface, 
where $W$ is the band-width of the electrons. The critical value of 
$K_{c} \sim \frac{V^2}{W}$ where this takes place
implies $T_{K}\gg K$, leading to a small vison gap energy ($\Delta_v \sim
\frac{K^3}{V^2}$, ({ see Appendix \ref{appendixA}  }).  This suggests a continuous zero-temperature phase
transition via the collapse of the vison gap at the quantum critical
point.

		\begin{figure}[b]
		\includegraphics[scale=.13]{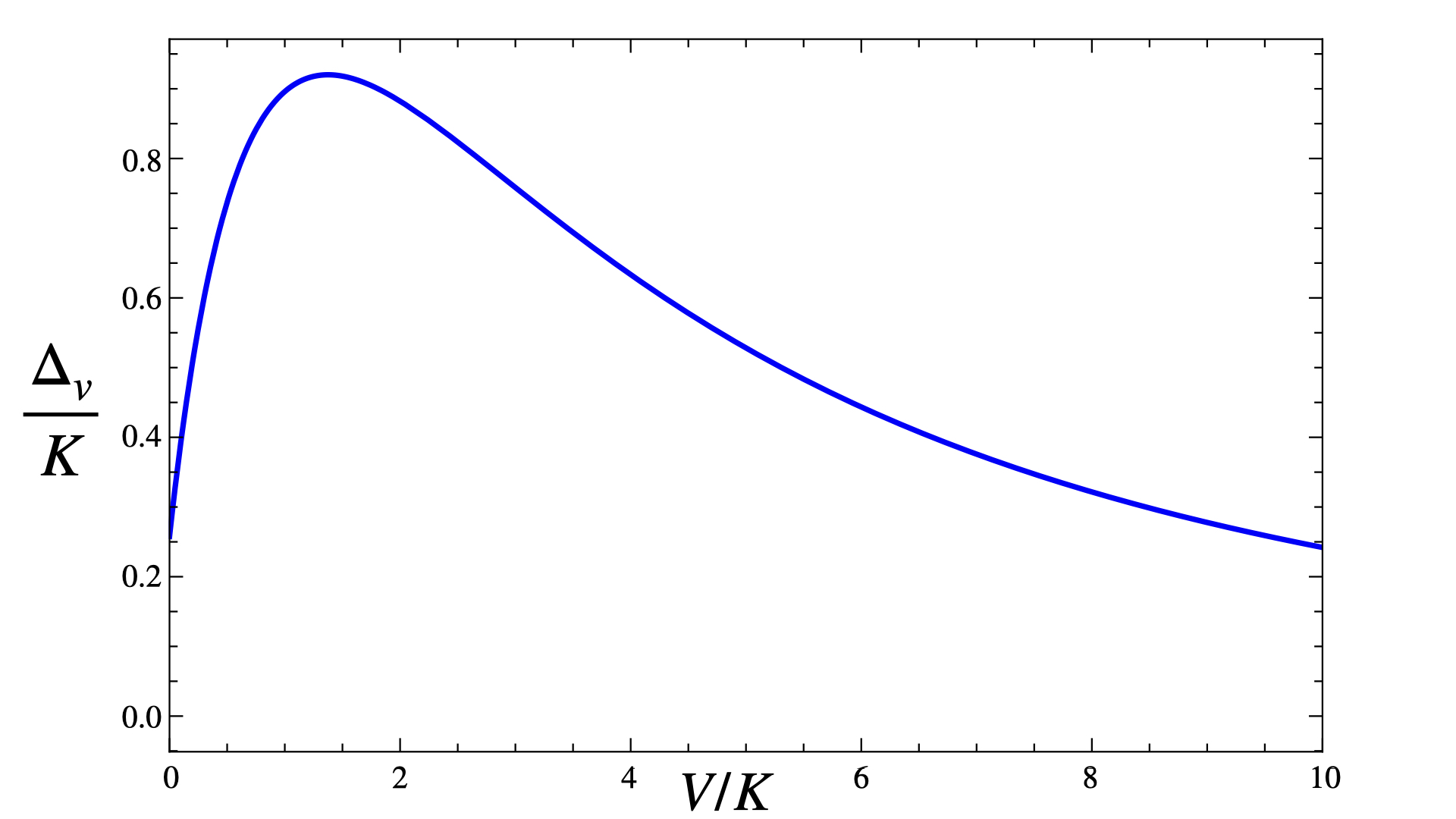}
		\caption{(a) The vison gap $\Delta_{v}$
associated with a single bond flip 
$\hat{u}_{ij}=1\rightarrow -1$. $\Delta_{v}$,
is the characteristic energy scale of $\mathbb{Z}_{2}$ gauge fluctuations in 
the CPT model. The figure shows the vison gap as a
function of the Kondo hybridization, calculated in the mean-field
theory of the CPT model: this scale 
forms the boundary of the Ising phase transition $T_{c1}$ in the regime $T_{K}>T_{c1}$. For large $J$, the hybridization $V\sim J$, so the large $J$ behavior of vison gap energy allows us to estimate the Ising temperature as a function of Kondo coupling $\Delta_v\sim T_{c1}\sim \frac{1}{J^2}$. }
		\label{visongap}
	\end{figure}

While the vison gap energy, $\Delta_v\sim T_{c1}\sim
\frac{1}{J^2}$\cite{eschmann_thermodynamic_2020} (Fig. \ref{visongap})
representing the mass gap of the $\mathbb{Z}_2$ gauge theory, remains
finite at all $J$ in mean-field calculations
({ Appendix \ref{appendixA}}), there is good reason to believe that
quantum fluctuations will suppress this gap to zero at a finite $J$, producing
a continuous superconducting-insulator transition. 
We expect that the  zero point energy of 
spin-wave fluctuations, $E_{Qfl}\sim J^{\alpha}$
is a monotonically increasing function of $J$ (i.e. $\alpha>0$). 
The spin-liquid 
will remain energetically stable 
against gauge fluctuations provided
$E_{Qfl}<\Delta_v$.  
As $J $ increases, $E_{Qfl}\sim J^{\alpha }$ rises, while
$\Delta_{v}\sim 1/J^{2}$ falls continuously, so we expect that 
this inequality fails at  a finite critical $J_{c}$. 
Beyond this point, mean-field theory breaks
down due to gauge fluctuations, and the system  will transition into a
Kondo insulator with a decoupled Kitaev orbital liquid.

In addition to the spin-wave fluctuations, 
$\mathbb{Z}_2$ gauge fluctuations will also tend to suppress the stability of 
the $1e$ superconductor.
From our calculations of the energy spectrum associated with a single
$\mathbb{Z}_{2}$ bond flip  ({ Appendix \ref{appendixA}}: Fig. \ref{phaseshift}), we observe the associated generation
of spin $S=1$ resonant
Majorana modes within the spin liquid. When
the superconducting gap  $V$ exceeds these resonant energies, they
sharpen into long-lived bound-states.
The concentration of these unscreened moments associated with
bond-flips will follow an 
activated temperature dependence, $N\sim
\exp(\frac{- (E_{bound}+\Delta_{v})}{T})$, where $E_{bound}$ is the Majorana
bound-state energy. 
The appearance of unscreened triplet states implies a
weakening of the Kondo screening associated with $\mathbb{Z}_{2}$ bond-flips in
the superconducting phase, ultimately making the Kondo insulator,
with its robust screening, more energetically favorable.

For these reasons, we believe that the Ising phase boundaries of the
$1e$ superconductor and the Kitaev orbital liquid will merge at a
quantum multicritical point. However, to definitively establish the
nature of this quantum critical point requires further numerical
investigation of the model using Monte Carlo and DMRG (on a lower
dimensional analog), etc.

We now discuss the nature of the $1e$ superconducting phase and its 
finite temperature superconducting phase boundary.  The $1e$
superconductor is a $\mathbb{Z}_{2}$ Higgs phase, and to understand its broken
symmetry and its finite temperature phase diagram,  we must go beyond mean-field theory. The mean-field 
spinor
order parameter 
\begin{equation}\label{}
V (x_{j})
 = \langle (\vec{\sigma }\cdot \vec{\chi }_{j})c_{j}\rangle 
\end{equation}
carries a $\mathbb{Z}_{2}$ gauge charge, and 
Elitzur's theorem guarantees that this quantity will average to zero
under the thermal $\mathbb{Z}_{2}$ gauge fluctuations. 
However, this does not
rule out the development of charge $2e$ composite triplet order associated with
\begin{equation}\label{}
\vec{\Psi}_{2e} = \langle  V^{T} (x) i \sigma_{2}\vec \sigma V (x)\rangle 
\end{equation}
which is $\mathbb{Z}_{2}$ gauge invariant, nor does it rule out the 
the possibility 
of gauged off-diagonal
long-range order, in which the charge $1e$
spinor order parameters at sites $j$ and $i$ 
are linked by a
product of gauge fields between the two sites,
\begin{equation}\label{}
\langle \hat V\dg (x)\hat {{\cal P}}(x,y)\hat V (y)\rangle   \xrightarrow[]{|x-y|\rightarrow
\infty} 
V\dg (x)V (y)
\end{equation}
where $\hat {{\cal P}} (x,y) = \prod_{l}u_{(l+1,l)}$ along a path from
$y$ to $x$. 

\begin{figure}[h]
	\includegraphics[scale=.13]{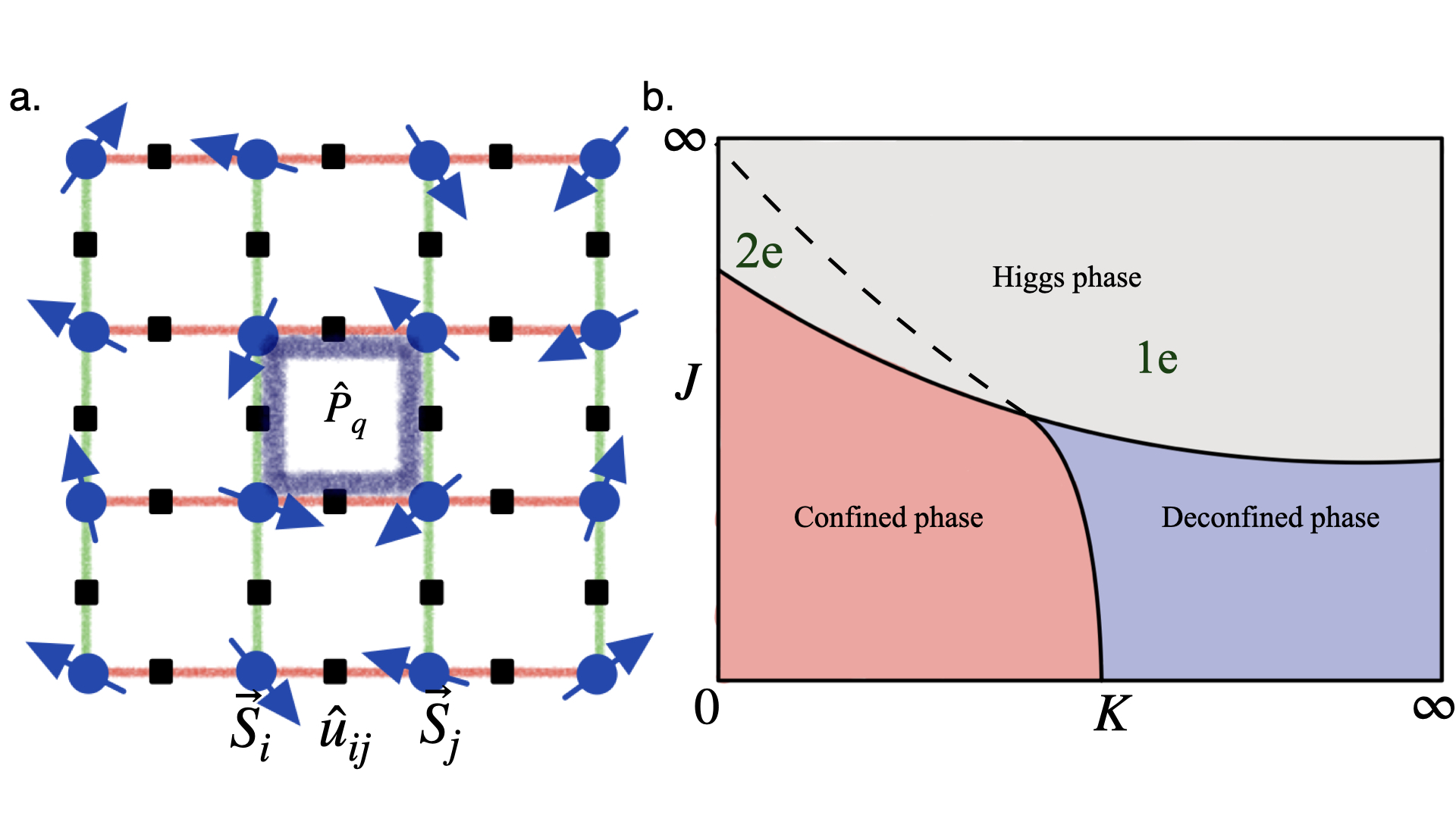} \caption{(a) Depicts a
		$\mathbb{Z}_2$ gauge XY model where the $XY$ spin interactions are
		gauged by $\mathbb{Z}_2$ bond variables $\hat{u}_{ij}$. Additionally,
		each plaquette $P_{q}$ is formed by the product of $\hat{u}_{ij}$
		around the square, costs an energy $K$ when going from $P_{q}=1$ to
		$P_{q}=-1$.  (b) Shows the phase diagram of $\mathbb{Z}_2$ gauged XY
		model in 3D cubic lattice. The model has 3 phases, 1. {\it Higgs
			phase}: In the Higgs phase, the spin $\vec{S}$ gains long range order
		and vacuum expectation value. The dashed line
		represents the speculated phase boundary between the
		long-range ordered 
$e^{i\theta }$ and $e^{2i \theta }$ Higgs phases.  2. {\it Confined phase}: The confined
		phase is marked by the presence of deconfined $P_{q}=-1$, $\pi$-flux
		plaquettes spread through the ground state. 3. {\it Deconfined phase}:
		In this phase, $\pi$ flux plaquettes are absent and $\vec{S}_i$s
		aren't long range ordered. } \label{quantumXY} \end{figure}

These two possibilities can be explored further by integrating out 
out the fermions and then expanding the effective action to leading
order in the hybridization and $\mathbb{Z}_{2}$ bond variables. The 
corresponding statistical mechanical model for the finite temperature
behavior is a $\mathbb{Z}_{2}$ gauged spinor model, 
\begin{equation}\label{}
	H_{SU (2)} = - J \sum_{\langle i,j\rangle } \biggl[
	{V}\dg_{i}
	u_{ij}V_{j}+ {\rm H. c}\biggr] - K \sum_{q }\prod_{\partial P_q}
	\hat{u}_{ij}.\small
\end{equation}
In the presence of magnetic anisotropy or a Zeeman splitting, we can
replace $V_{j}$
by an x-y order parameter, $V_{j}\rightarrow e^{i \theta_{j}}$ ,
so that this model reduces to the 
3D $\mathbb{Z}_{2}$-gauged $XY$
model (or its 2+1 dimensional quantum equivalent)
\cite{lammert_topology_1993,lammert_topology_1995,sachdev_topological_2019},
where a lattice of $XY$ rotors is gauged by $\mathbb{Z}_2$ field (Fig
\ref{quantumXY}). The Hamiltonian for the $\mathbb{Z}_2$ gauged
$XY$ model is given by: \begin{equation}
	H_{qXY}=-J\sum_{\langle i j \rangle}
	{u}_{ij}\cos(\theta_i-\theta_j)-K\sum_{q} 
	\prod_{\partial P_q}
	\hat{u}_{ij}\small
	\label{Z2XY} \end{equation}
Here, each site has an $XY$ order parameter parameterized by
$\theta_i$, and the bonds have a $\mathbb{Z}_2$ gauge field
$\hat{u}_{ij}$ on them (Fig. \ref{quantumXY}a), which multiplies the
matter field $\theta_i$s. Each flux plaquette $P_q$ (``vison'') at $q$
(Fig. \ref{quantumXY}a), costs an energy $2K$ in the 3D cubic
lattice.  At small $J$, this model undergoes a pure gauge transition at a finite critical $K_c$ in which the visons become linearly confined.

A more nuanced analysis of the 3D gauged $XY$ model (\ref{Z2XY}) is
necessitated to understand the Higgs phase of the
model.  The J-K phase diagram of this model is understood in the
various limits $J=0$, $K=0$, $K=\infty $ and $J=\infty $.
At $J=0$, there is a deconfinement transition at a finite $K=K_{c}$
into a phase where visons become linearly confined.  At $K=0$ there is an x-y phase transition into a phase where $\Psi_{2e }=
e^{2i \theta}$ develops long range order, whilst at $K=\infty $, where the
$\mathbb{Z}_{2}$ degrees of freedom are quenched, there is an x-y transition
into a phase where $e^{i\theta }$ develops long range order.  Finally at
$J=\infty $, the ground-state manifold where
$\cos(\theta_i-\theta_j)=u_{ij}$, is identical to that of the $\mathbb{Z}_{2}$
gauged Ising model considered by Fradkin and Shenkar\cite{fradkin_phase_1979}, which has no
phase boundary.

What is not known about the $\mathbb{Z}_{2}$ gauged Ising model, is whether
the Ising deconfinement transition continues into the ordered phase.
If the continuation is present, then there would be a vison-confining 
phase transition between a charge $2e$ and charge $1e$ order parameter.
The phase boundary of this transition would have to continue to one of
the corners of the phase diagram, that is to $J=\infty $ and either
$K=0$ or $K=\infty $ (see Fig \ref{quantumXY}b).

In drawing our tentative phase diagram of the CPT model, we have
assumed that no such phase boundary exists. However, if it does, then the
Ising transition, $T_{c1}$  would extend inside the
superconducting phase,  corresponding to a transition from a charge-$e$ fractionalized ordered
phase to a charge-$2e$ composite ordered
phase\cite{komijani_order_2018}. Future numerical
analysis is required to confirm the existence of this transition.

One of the intriguing features of the proposed phase diagram
(Fig. \ref{PhaseDiagram}) are the 
two routes into the order-fractionalized phase: one originating from an FL$^{*}$ phase intertwined with a decoupled spin
liquid, the other emerging directly from a heavy Fermi liquid.
These two routes are reminiscent of the Bose-Einstein and  BCS
condensation pathways to a conventional superconductor.
This raises
the fascinating possibility that spinor superconducting order might develop
as a novel superconducting instability of a Landau Fermi liquid.
Unlike conventional triplet order, the fractionalized 
spinor order exhibits Kramer's
degeneracy and  a spontaneous
broken time-reversal symmetry, even in situations where there are no
conventional two-dimensional triplet representations, such as the orthorhombic
triplet superconductor UTe$_{2}$. At present, the issue of whether UTe$_{2}$
spontaneously breaks time-reversal symmetry is a controversial
point. Were the ground-state of this novel material to support broken
time-reversal symmetry via a single sharp phase transition, 
this would constitute evidence for a fractionalized superconducting order
parameter.


	\begin{acknowledgments}
		\textit{Acknowledgments:}
			This work was supported by Office of Basic Energy Sciences, Material
		Sciences and Engineering Division, U.S. Department of Energy (DOE)
		under Contracts No. DE-SC0012704 (AMT) and DE-FG02-99ER45790 (AP and PC ).
	\end{acknowledgments}

	\appendix
	\begin{widetext}
			\section{Vison gap energy and Ising phase transition}
			\label{appendixA}
	\end{widetext}
	
	In the CPT model \cite{coleman_solvable_2022,tsvelik_order_2022}, the bond variables $\hat{u}_{ij}$ commute with Kondo interaction and remain constants of motion. Consequently, the gauge fluctuations in the CPT model are thermal. Thermal gauge fluctuations in the hyper-octagonal Yao Lee spin liquid subside below the Ising critical temperature $T_{c,Ising}$.  Free majoranas are the low-energy excitations in this regime. 
	
	In the order-fractionalized phase, the condensation of
	electron majorana pairs gaps out the majorana spectrum and
	Higgses the $\mathbb{Z}_2$ and Maxwell fields. This impacts
	the vison gap energy $\Delta_v$, (the energy cost of one bond
	flip) in the fractionalized ordered phase. Given that the
	vison gap energy is the characteristic energy of
	$\mathbb{Z}_2$ gauge, the Ising critical temperature 
varies as the vison gap energy $T_{c1}\sim\Delta_v$.  Indeed, 
Monte-Carlo simulations
\cite{eschmann_thermodynamic_2020,obrien_classification_2016} show
that in  3D Kitaev spin liquids, the Ising critical temperature
$T_{c,Ising}$, and the vison gap energy $\Delta_v$ are linearly
correlated. Thus estimating vison gap energy $\Delta_v$ provides an
estimate for the Ising critical temperature. 
	
	\subsection{Vison gap energy of CPT Model} 
	
\begin{figure}[h]
\includegraphics[scale=.12]{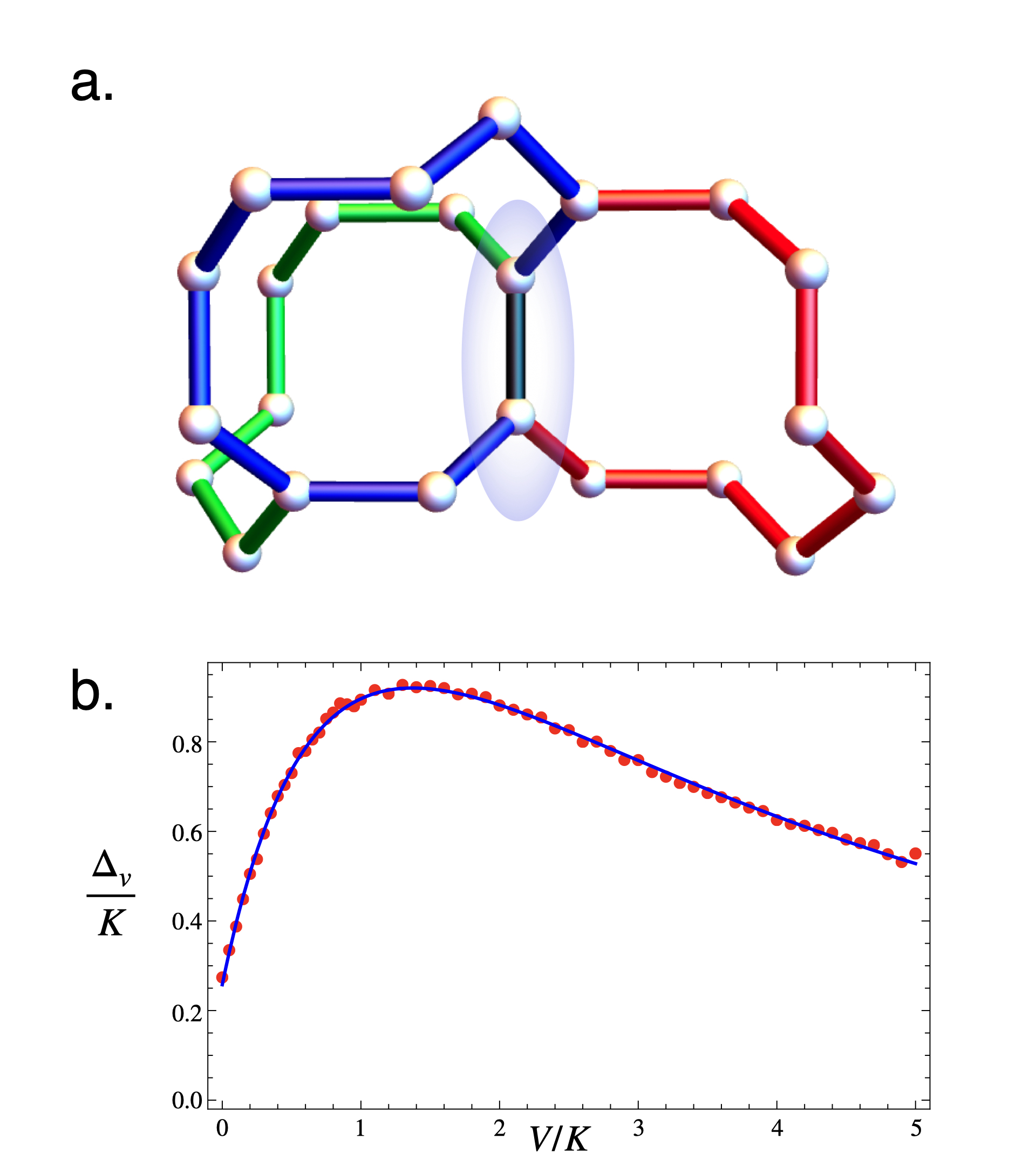} \caption{(a) Shows the
plaquettes in hyperoctagon lattice that change signs when one flips a
bond variable $\hat{u}_{ij}=1\rightarrow -1$ shown in the oval, the
energy. The energy cost associated with this flip is the vison gap
energy (b)Vison gap energy as a function of hybridization calculated
numerically on a 50x50x50 unit cell lattice, with a hybridization
resolution of $\frac{\Delta V}{K}=0.01$. Consequently, we find the
best fit for the vison gap energy as a function of hybridization
$V$. } \label{bondflip} \end{figure}

One can analytically determine the vison gap energy for Kitaev-like
spin-liquids \cite{panigrahi_analytic_2023} by evaluating the change
in free energy associated with flipping a local $\mathbb{Z}_2$
variable $\hat{u}_{ij}$ (Fig. \ref{bondflip}) away from the ground
state gauge configuration. A flip in the local $\mathbb{Z}_2$ variable
$\hat{u}_{ij}$ acts as an impurity potential, leading to a scattering
phase shift due to the Majoranas and a change in free energy, i.e.,
the vison gap energy $\Delta_v$. This approach is validated against
vison gap energy values obtained via Monte Carlo simulations,
particularly for the $V=0$ limit. For the isotropic case
$J_x=J_y=J_z=K$, Monte Carlo simulations yield a vison gap energy of
$\Delta_v=0.09(1)K$\cite{obrien_classification_2016} for the Kitaev spin
liquid on a hyperoctagon lattice (i.e. (10,3)a system). Our
calculation aligns precisely with this result, yielding a vison gap
energy of $\Delta_v=0.089(5)K$ per Majorana species $\chi^a$ i.e. the
Yao-Lee equivalent of Kitaev spin liquid. This methodology remains
robust even when the Majoranas hybridize with electrons, allowing us
to extend the approach to the CPT model.  

	To compute the vison gap energy in the CPT model, we flip the $\hat{u}_{ij}$ bond between the $2^{nd}$ and $3^{rd}$ atoms in the unit cell positioned at the origin,

	\begin{equation}
		H_{CPT+3v}=H_{CPT}-2 K(i \vec{\chi}_{0,2}\cdot\vec{\chi}_{0,3}).
		\label{BondImpurityHam}
	\end{equation} 
	
	$H_{CPT+3v}$ is the Hamiltonian that has a local $\mathbb{Z}_2$ bond flip over this ground-state gauge configuration. This bond flip is associated with the creation of 3 visons adjacent to the bond. Treating the bond-flip term as an impurity potential,
	
	\begin{equation}
		\hat{V}_{flip}=-2 K(i \vec{\chi}_{0,2}\cdot\vec{\chi}_{0,3})
	\end{equation}
	allows us to calculate the associated free-energy change. The free energy of the CPT model with $\mathbb{Z}_2$ bond flip is expressed in terms of its bare-Green's function $G_{0,CPT}$,
	
	\begin{equation}
		\beta F=-\frac{1}{2}Tr\log[-G_{0,CPT}^{-1}+\hat{V}_{flip}]
	\end{equation}
	where $G_{0,CPT}(\omega,\vec{k})=(\omega \mathbb{I}-h_{CPT}(\vec{k}) )^{-1} $ is the Green's function for the CPT Model in the mean-field configuration.
	
	Since the bond-flip potential $\hat{V}_{flip}$ scatters Majorana fermions in the Yao-Lee spin liquid. The associated free-energy change is obtained in terms of effective majorana green's function $G_{\vec{\chi}}$, which includes self-energy corrections from the electron-majorana condensate, as follows
	
	\begin{equation}
		\Delta F=\frac{1}{2\beta}Tr\log[1-\hat{V}_{flip}G_{\vec{\chi}}].
	\end{equation}
	Here, the trace is over the system and Matsubara frequencies. 
	
	The effective Majorana Green's function in the Majorana-electron condensate is given by,
	\begin{equation}
		\large	G_{\vec{\chi}}(z,\vec{k},V)=\frac{1}{z-K h(\vec{k})-\frac{V^2}{z+t\, h(\vec{k})}}
	\end{equation}
	Where $V$ is the magnitude of the spinor order, and $h(\vec{k})$ is the $4$-band Hamiltonian given in equation (\ref{YLHam}). This effective Green's function is used to calculate vison gap energy $\Delta_v$ by calculating the scattering phase shift of the $\mathbb{Z}_2$ bond-flip potential.
	
	The scattering potential $\hat{V}_{flip}$ is local, the free-energy of $\mathbb{Z}_2$ bond flip potential can be re-expressed in terms of the local majorana Green's function $g_{\vec{\chi}}(i\omega_n)$ 
	\begin{equation}
		\Delta F=\frac{1}{2\beta}\sum_{i\omega_n}[\ln(1-\hat{V}_{flip}g_{\vec{\chi}}(i\omega_n))]
	\end{equation}
	where, the local majorana Green's function  $g_{\vec{\chi}}(i\omega_n)$ is,
	\begin{equation}
		g_{\vec{\chi}}(z)=\frac{1}{N_c}\sum_{k\in BZ}G_{\vec{\chi}}(z,k)
	\end{equation}
	obtained by summing the Majorana Green's function over the Brillouin zone. 
		\begin{figure}[h]
		\includegraphics[scale=.14]{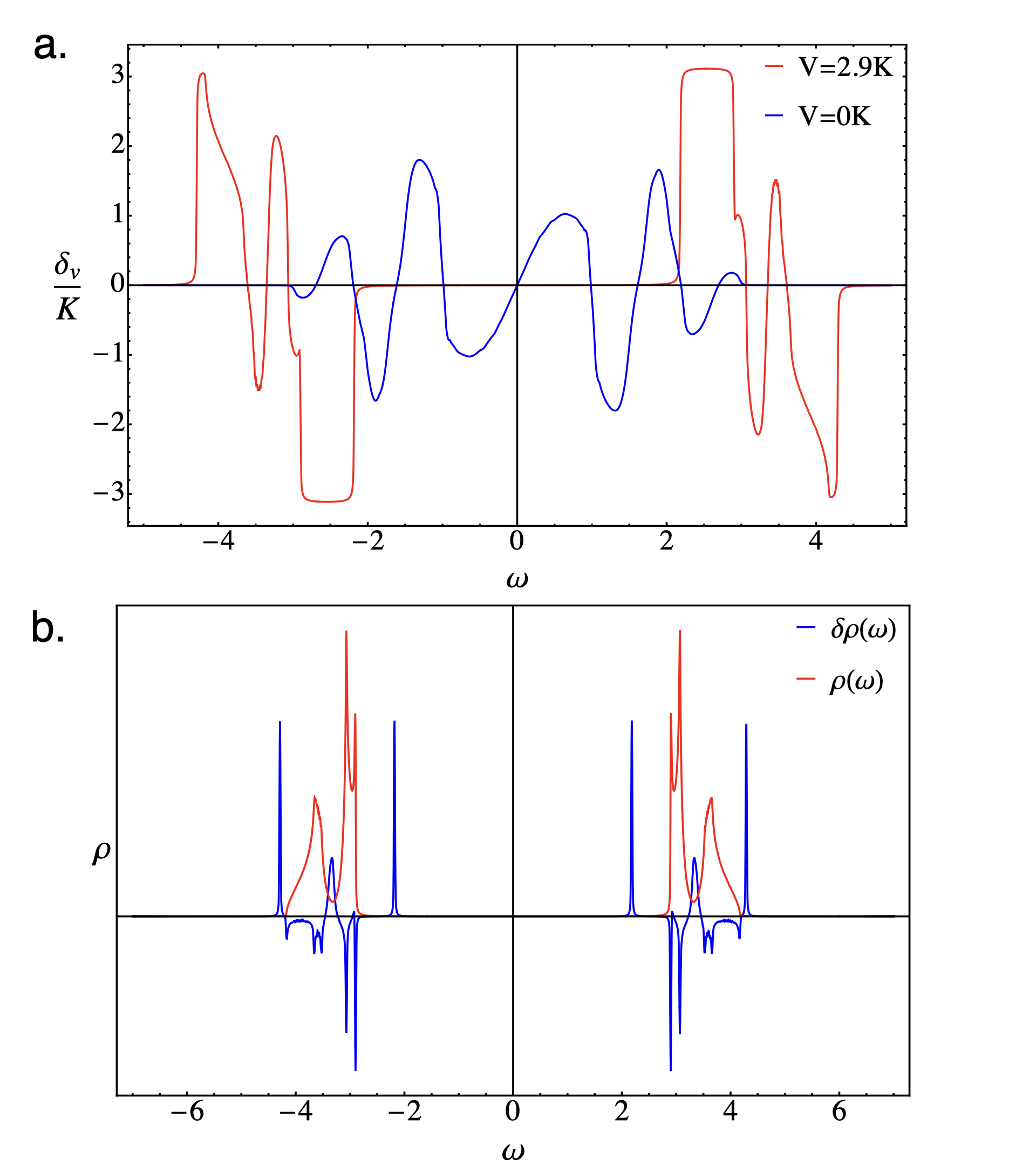}
		\caption{(a) 
				Depicts the scattering phase shift associated with a bond flip impurity $\hat{V}_{flip}$ for a species of spin Majorana $\chi^a$, illustrating two spinor order strengths: $V=0 K$ (in blue) and $V=2.9 K$ (in red). The plateau for the gapped $V=2.9K$ state at $\Delta_v=\pi$ signifies the formation of an in-gap bound state.(b)Shows the change in density of states $\delta \rho=\frac{1}{2\pi}\frac{d \delta_v(\omega)}{d\omega}$ (in blue) associated with the bond flip potential $\hat{V}_{flip}$. The sharp peak inside the gap (shown using the bulk density of state $\rho$ for $V=2.9 K$ in red) signifies the presence of an in-gap bound state for spinor order strength of $V=2.9 K$. }
		\label{phaseshift}
	\end{figure} 
	Upon carrying out the Matsubara frequency summation, one obtains the expression of the free-energy change in terms of the scattering phase shift,  
	\begin{equation}
		\Delta F =\int_{-\infty}^{\infty}\frac{d\omega}{2\pi}\left(\frac{1}{2}-f(\omega)\right)\delta_v(\omega)
	\end{equation}
	where, the scattering phase shift $\delta_v$ is given by
	\begin{equation}
		\delta_v(\omega)= Im \tr\log[1-\hat{V}_{flip}g_{\vec{\chi}}(z)]_{z=\omega-i\delta}.
	\end{equation}
	At zero-temperature, this free-energy change corresponds to the vison gap energy $\Delta_v$ given by,
	\begin{equation}
		\Delta_v=-K\int_{-\infty}^{0}\frac{dx}{2\pi}Im \log[\det(1-\hat{V}_{flip}g_{\vec{\chi}}(z))]_{z=x-i\delta}.
		\label{visonE}
	\end{equation}
	Where the $\tr$ is over the sites within the unit-cells. Numerically, the vison gap energy calculation was carried out on a $50\times 50\times 50$ lattice by carrying out discrete summation over the momentum in the Brillouin zone and frequencies. 
	
	We compute the vison gap energy for discrete values of hybridization $V$ with a resolution of $\frac{\Delta V}{t}=0.01$. The resulting dataset was fit to obtain a functional form,
	\begin{equation}
		\Delta_v(V)=	r\left(\frac{1}{(s V+u)^2}-\frac{1}{(s V + u)^4}\right)
	\end{equation}
	with $s=0.273$, $u=1.040$ and $r=3.680$. This function form matches the asymptotic behavior of the vison gap energy $\Delta_v\sim \frac{1}{V^2}$ for large $V$, which is always positive and is obtained analytically using equation (\ref{visonE}). Additionally, it also matches the numerical result \cite{obrien_classification_2016} for $V=0$ as can be seen in (Fig. \ref{bondflip}), and consequently is a reliable fit. The initial enhancement in the vison gap energy $\Delta_v$ is a consequence of $\mathbb{Z}_2$ gauge theories being Higgs phases of continuum gauge theories \cite{fradkin_quantum_2021,sachdev_topological_2019}, thus electron-majorana fractionalized order formation enhances the already massive $\mathbb{Z}_2$ gauge fields. The reduction in the vison gap energy $\Delta_v$ for large hybridization $\Delta_v$ is a result of renormalization effects at large Kondo coupling $J$.   
	
	Since the vison gap energy in the CPT model asymptotically decreases as $\Delta_v \sim \frac{1}{V^2}$ for large hybridization $V$.  Thus, as the Kondo coupling $J$ increases, the characteristic vison gap energy scale renormalizes to smaller values as a result of Kondo screening. Additionally, given the linear correlation between vison-gap energy and Ising critical temperature $T_{c,Ising}$, we estimate that Ising critical temperature $T_{c,Ising}\sim \Delta_v\sim \frac{1}{V^2}$ reduces with increasing Kondo coupling. Beyond this Ising critical temperature scale $T_{c,Ising}$, the thermal gauge fluctuations destroy electron-majorana fractionalized order. 
	
	The electron-majorana fractionalized order in the CPT model is
expected to be suppressed by quantum fluctuations about the
mean-field theory. Such quantum fluctuations will grow 
as one moves away fromthe small Kondo coupling $J$ regime,
where the $J$ is the small parameter that controlling the
mean-field treatment. In the large Kondo coupling $J$ limit, the
ground state is a Kondo insulator with a decoupled orbital Kitaev spin
liquid. Thus, passing from the superconducting 
 electron-majorana condensate at small $J$ to the Kondo insulator
 phase at large $J$,  the system
undergoes a quantum phase transition. Away from half-filling, this
quantum phase transition is associated with the small-to-large expansion of the
neutral Fermi surface, a likely signature of a continuous 
quantum phase transition.


\begin{thebibliography}{0}%
\makeatletter
\providecommand \@ifxundefined [1]{%
 \@ifx{#1\undefined}
}%
\providecommand \@ifnum [1]{%
 \ifnum #1\expandafter \@firstoftwo
 \else \expandafter \@secondoftwo
 \fi
}%
\providecommand \@ifx [1]{%
 \ifx #1\expandafter \@firstoftwo
 \else \expandafter \@secondoftwo
 \fi
}%
\providecommand \natexlab [1]{#1}%
\providecommand \enquote  [1]{``#1''}%
\providecommand \bibnamefont  [1]{#1}%
\providecommand \bibfnamefont [1]{#1}%
\providecommand \citenamefont [1]{#1}%
\providecommand \href@noop [0]{\@secondoftwo}%
\providecommand \href [0]{\begingroup \@sanitize@url \@href}%
\providecommand \@href[1]{\@@startlink{#1}\@@href}%
\providecommand \@@href[1]{\endgroup#1\@@endlink}%
\providecommand \@sanitize@url [0]{\catcode `\\12\catcode `\$12\catcode
  `\&12\catcode `\#12\catcode `\^12\catcode `\_12\catcode `\%12\relax}%
\providecommand \@@startlink[1]{}%
\providecommand \@@endlink[0]{}%
\providecommand \url  [0]{\begingroup\@sanitize@url \@url }%
\providecommand \@url [1]{\endgroup\@href {#1}{\urlprefix }}%
\providecommand \urlprefix  [0]{URL }%
\providecommand \Eprint [0]{\href }%
\providecommand \doibase [0]{https://doi.org/}%
\providecommand \selectlanguage [0]{\@gobble}%
\providecommand \bibinfo  [0]{\@secondoftwo}%
\providecommand \bibfield  [0]{\@secondoftwo}%
\providecommand \translation [1]{[#1]}%
\providecommand \BibitemOpen [0]{}%
\providecommand \bibitemStop [0]{}%
\providecommand \bibitemNoStop [0]{.\EOS\space}%
\providecommand \EOS [0]{\spacefactor3000\relax}%
\providecommand \BibitemShut  [1]{\csname bibitem#1\endcsname}%
\let\auto@bib@innerbib\@empty
\end{thebibliography}%


\begin{thebibliography}{35}%
	\makeatletter
	\providecommand \@ifxundefined [1]{%
		\@ifx{#1\undefined}
	}%
	\providecommand \@ifnum [1]{%
		\ifnum #1\expandafter \@firstoftwo
		\else \expandafter \@secondoftwo
		\fi
	}%
	\providecommand \@ifx [1]{%
		\ifx #1\expandafter \@firstoftwo
		\else \expandafter \@secondoftwo
		\fi
	}%
	\providecommand \natexlab [1]{#1}%
	\providecommand \enquote  [1]{``#1''}%
	\providecommand \bibnamefont  [1]{#1}%
	\providecommand \bibfnamefont [1]{#1}%
	\providecommand \citenamefont [1]{#1}%
	\providecommand \href@noop [0]{\@secondoftwo}%
	\providecommand \href [0]{\begingroup \@sanitize@url \@href}%
	\providecommand \@href[1]{\@@startlink{#1}\@@href}%
	\providecommand \@@href[1]{\endgroup#1\@@endlink}%
	\providecommand \@sanitize@url [0]{\catcode `\\12\catcode `\$12\catcode
		`\&12\catcode `\#12\catcode `\^12\catcode `\_12\catcode `\%12\relax}%
	\providecommand \@@startlink[1]{}%
	\providecommand \@@endlink[0]{}%
	\providecommand \url  [0]{\begingroup\@sanitize@url \@url }%
	\providecommand \@url [1]{\endgroup\@href {#1}{\urlprefix }}%
	\providecommand \urlprefix  [0]{URL }%
	\providecommand \Eprint [0]{\href }%
	\providecommand \doibase [0]{https://doi.org/}%
	\providecommand \selectlanguage [0]{\@gobble}%
	\providecommand \bibinfo  [0]{\@secondoftwo}%
	\providecommand \bibfield  [0]{\@secondoftwo}%
	\providecommand \translation [1]{[#1]}%
	\providecommand \BibitemOpen [0]{}%
	\providecommand \bibitemStop [0]{}%
	\providecommand \bibitemNoStop [0]{.\EOS\space}%
	\providecommand \EOS [0]{\spacefactor3000\relax}%
	\providecommand \BibitemShut  [1]{\csname bibitem#1\endcsname}%
	\let\auto@bib@innerbib\@empty
	\bibitem [{\citenamefont {Petrovic}\ \emph {et~al.}(2001)\citenamefont
		{Petrovic}, \citenamefont {Pagliuso}, \citenamefont {Hundley}, \citenamefont
		{Movshovich}, \citenamefont {Sarrao}, \citenamefont {Thompson}, \citenamefont
		{Fisk},\ and\ \citenamefont {Monthoux}}]{petrovic_heavy-fermion_2001}%
	\BibitemOpen
	\bibfield  {author} {\bibinfo {author} {\bibfnamefont {C.}~\bibnamefont
			{Petrovic}}, \bibinfo {author} {\bibfnamefont {P.~G.}\ \bibnamefont
			{Pagliuso}}, \bibinfo {author} {\bibfnamefont {M.~F.}\ \bibnamefont
			{Hundley}}, \bibinfo {author} {\bibfnamefont {R.}~\bibnamefont {Movshovich}},
		\bibinfo {author} {\bibfnamefont {J.~L.}\ \bibnamefont {Sarrao}}, \bibinfo
		{author} {\bibfnamefont {J.~D.}\ \bibnamefont {Thompson}}, \bibinfo {author}
		{\bibfnamefont {Z.}~\bibnamefont {Fisk}},\ and\ \bibinfo {author}
		{\bibfnamefont {P.}~\bibnamefont {Monthoux}},\ }\bibfield  {title} {\bibinfo
		{title} {Heavy-fermion superconductivity in {CeCoIn} $_{\textrm{5}}$ at 2.3
			{K}},\ }\href {https://doi.org/10.1088/0953-8984/13/17/103} {\bibfield
		{journal} {\bibinfo  {journal} {Journal of Physics: Condensed Matter}\
		}\textbf {\bibinfo {volume} {13}},\ \bibinfo {pages} {L337} (\bibinfo {year}
		{2001})}\BibitemShut {NoStop}%
	\bibitem [{\citenamefont {Flint}\ \emph {et~al.}(2008)\citenamefont {Flint},
		\citenamefont {Dzero},\ and\ \citenamefont {Coleman}}]{flint_heavy_2008}%
	\BibitemOpen
	\bibfield  {author} {\bibinfo {author} {\bibfnamefont {R.}~\bibnamefont
			{Flint}}, \bibinfo {author} {\bibfnamefont {M.}~\bibnamefont {Dzero}},\ and\
		\bibinfo {author} {\bibfnamefont {P.}~\bibnamefont {Coleman}},\ }\bibfield
	{title} {{\bibinfo {title} {Heavy electrons and the
				symplectic symmetry of spin}},\ } \href {https://doi.org/10.1038/nphys1024}
	{\bibfield  {journal} {\bibinfo  {journal} {Nature Physics}\ }\textbf
		{\bibinfo {volume} {4}},\ \bibinfo {pages} {643} (\bibinfo {year}
		{2008})}\BibitemShut {NoStop}%
	\bibitem [{\citenamefont {Tsvelik}\ and\ \citenamefont
		{Coleman}(2022)}]{tsvelik_order_2022}%
	\BibitemOpen
	\bibfield  {author} {\bibinfo {author} {\bibfnamefont {A.~M.}\ \bibnamefont
			{Tsvelik}}\ and\ \bibinfo {author} {\bibfnamefont {P.}~\bibnamefont
			{Coleman}},\ }\bibfield  {title} {{\bibinfo {title}
			{Order fractionalization in a {Kitaev}-{Kondo} model}},\ }\href
	{https://doi.org/10.1103/PhysRevB.106.125144} {\bibfield  {journal} {\bibinfo
			{journal} {Physical Review B}\ }\textbf {\bibinfo {volume} {106}},\ \bibinfo
		{pages} {125144} (\bibinfo {year} {2022})}\BibitemShut {NoStop}%
	\bibitem [{\citenamefont {Coleman}\ \emph {et~al.}(2022)\citenamefont
		{Coleman}, \citenamefont {Panigrahi},\ and\ \citenamefont
		{Tsvelik}}]{coleman_solvable_2022}%
	\BibitemOpen
	\bibfield  {author} {\bibinfo {author} {\bibfnamefont {P.}~\bibnamefont
			{Coleman}}, \bibinfo {author} {\bibfnamefont {A.}~\bibnamefont {Panigrahi}},\
		and\ \bibinfo {author} {\bibfnamefont {A.}~\bibnamefont {Tsvelik}},\
	}\bibfield  {title} {{\bibinfo {title} {Solvable {3D}
				{Kondo} {Lattice} {Exhibiting} {Pair} {Density} {Wave}, {Odd}-{Frequency}
				{Pairing}, and {Order} {Fractionalization}}},\ }\href
	{https://doi.org/10.1103/PhysRevLett.129.177601} {\bibfield  {journal}
		{\bibinfo  {journal} {Physical Review Letters}\ }\textbf {\bibinfo {volume}
			{129}},\ \bibinfo {pages} {177601} (\bibinfo {year} {2022})}\BibitemShut
	{NoStop}%
	\bibitem [{\citenamefont {Choi}\ \emph {et~al.}(2018)\citenamefont {Choi},
		\citenamefont {Klein}, \citenamefont {Rosch},\ and\ \citenamefont
		{Kim}}]{choi_topological_2018}%
	\BibitemOpen
	\bibfield  {author} {\bibinfo {author} {\bibfnamefont {W.}~\bibnamefont
			{Choi}}, \bibinfo {author} {\bibfnamefont {P.~W.}\ \bibnamefont {Klein}},
		\bibinfo {author} {\bibfnamefont {A.}~\bibnamefont {Rosch}},\ and\ \bibinfo
		{author} {\bibfnamefont {Y.~B.}\ \bibnamefont {Kim}},\ }\bibfield  {title}
	{{\bibinfo {title} {Topological superconductivity in the
				{Kondo}-{Kitaev} model}},\ }\href
	{https://doi.org/10.1103/PhysRevB.98.155123} {\bibfield  {journal} {\bibinfo
			{journal} {Physical Review B}\ }\textbf {\bibinfo {volume} {98}},\ \bibinfo
		{pages} {155123} (\bibinfo {year} {2018})}\BibitemShut {NoStop}%
	\bibitem [{\citenamefont {Seifert}\ \emph {et~al.}(2018)\citenamefont
		{Seifert}, \citenamefont {Meng},\ and\ \citenamefont
		{Vojta}}]{seifert_fractionalized_2018}%
	\BibitemOpen
	\bibfield  {author} {\bibinfo {author} {\bibfnamefont {U.~F.~P.}\
			\bibnamefont {Seifert}}, \bibinfo {author} {\bibfnamefont {T.}~\bibnamefont
			{Meng}},\ and\ \bibinfo {author} {\bibfnamefont {M.}~\bibnamefont {Vojta}},\
	}\bibfield  {title} {{\bibinfo {title} {Fractionalized
				{Fermi} liquids and exotic superconductivity in the {Kitaev}-{Kondo}
				lattice}},\ }\href {https://doi.org/10.1103/PhysRevB.97.085118} {\bibfield
		{journal} {\bibinfo  {journal} {Physical Review B}\ }\textbf {\bibinfo
			{volume} {97}},\ \bibinfo {pages} {085118} (\bibinfo {year}
		{2018})}\BibitemShut {NoStop}%
	\bibitem [{\citenamefont {Dzero}\ \emph {et~al.}(2010)\citenamefont {Dzero},
		\citenamefont {Sun}, \citenamefont {Galitski},\ and\ \citenamefont
		{Coleman}}]{dzero_topological_2010}%
	\BibitemOpen
	\bibfield  {author} {\bibinfo {author} {\bibfnamefont {M.}~\bibnamefont
			{Dzero}}, \bibinfo {author} {\bibfnamefont {K.}~\bibnamefont {Sun}}, \bibinfo
		{author} {\bibfnamefont {V.}~\bibnamefont {Galitski}},\ and\ \bibinfo
		{author} {\bibfnamefont {P.}~\bibnamefont {Coleman}},\ }\bibfield  {title}
	{{\bibinfo {title} {Topological {Kondo} {Insulators}}},\
	}\href {https://doi.org/10.1103/PhysRevLett.104.106408} {\bibfield  {journal}
		{\bibinfo  {journal} {Physical Review Letters}\ }\textbf {\bibinfo {volume}
			{104}},\ \bibinfo {pages} {106408} (\bibinfo {year} {2010})}\BibitemShut
	{NoStop}%
	\bibitem [{\citenamefont {Neupane}\ \emph {et~al.}(2013)\citenamefont
		{Neupane}, \citenamefont {Alidoust}, \citenamefont {Xu}, \citenamefont
		{Kondo}, \citenamefont {Ishida}, \citenamefont {Kim}, \citenamefont {Liu},
		\citenamefont {Belopolski}, \citenamefont {Jo}, \citenamefont {Chang},
		\citenamefont {Jeng}, \citenamefont {Durakiewicz}, \citenamefont {Balicas},
		\citenamefont {Lin}, \citenamefont {Bansil}, \citenamefont {Shin},
		\citenamefont {Fisk},\ and\ \citenamefont {Hasan}}]{neupane_surface_2013}%
	\BibitemOpen
	\bibfield  {author} {\bibinfo {author} {\bibfnamefont {M.}~\bibnamefont
			{Neupane}}, \bibinfo {author} {\bibfnamefont {N.}~\bibnamefont {Alidoust}},
		\bibinfo {author} {\bibfnamefont {S.-Y.}\ \bibnamefont {Xu}}, \bibinfo
		{author} {\bibfnamefont {T.}~\bibnamefont {Kondo}}, \bibinfo {author}
		{\bibfnamefont {Y.}~\bibnamefont {Ishida}}, \bibinfo {author} {\bibfnamefont
			{D.~J.}\ \bibnamefont {Kim}}, \bibinfo {author} {\bibfnamefont
			{C.}~\bibnamefont {Liu}}, \bibinfo {author} {\bibfnamefont {I.}~\bibnamefont
			{Belopolski}}, \bibinfo {author} {\bibfnamefont {Y.~J.}\ \bibnamefont {Jo}},
		\bibinfo {author} {\bibfnamefont {T.-R.}\ \bibnamefont {Chang}}, \bibinfo
		{author} {\bibfnamefont {H.-T.}\ \bibnamefont {Jeng}}, \bibinfo {author}
		{\bibfnamefont {T.}~\bibnamefont {Durakiewicz}}, \bibinfo {author}
		{\bibfnamefont {L.}~\bibnamefont {Balicas}}, \bibinfo {author} {\bibfnamefont
			{H.}~\bibnamefont {Lin}}, \bibinfo {author} {\bibfnamefont {A.}~\bibnamefont
			{Bansil}}, \bibinfo {author} {\bibfnamefont {S.}~\bibnamefont {Shin}},
		\bibinfo {author} {\bibfnamefont {Z.}~\bibnamefont {Fisk}},\ and\ \bibinfo
		{author} {\bibfnamefont {M.~Z.}\ \bibnamefont {Hasan}},\ }\bibfield  {title}
	{{\bibinfo {title} {Surface electronic structure of the
				topological {Kondo}-insulator candidate correlated electron system {SmB6}}},\
	}\href {https://doi.org/10.1038/ncomms3991} {\bibfield  {journal} {\bibinfo
			{journal} {Nature Communications}\ }\textbf {\bibinfo {volume} {4}},\
		\bibinfo {pages} {2991} (\bibinfo {year} {2013})}\BibitemShut {NoStop}%
	\bibitem [{\citenamefont {Song}\ and\ \citenamefont
		{Bernevig}(2022)}]{song_magic-angle_2022}%
	\BibitemOpen
	\bibfield  {author} {\bibinfo {author} {\bibfnamefont {Z.-D.}\ \bibnamefont
			{Song}}\ and\ \bibinfo {author} {\bibfnamefont {B.~A.}\ \bibnamefont
			{Bernevig}},\ }\bibfield  {title} {{\bibinfo {title}
			{Magic-{Angle} {Twisted} {Bilayer} {Graphene} as a {Topological} {Heavy}
				{Fermion} {Problem}}},\ }\href
	{https://doi.org/10.1103/PhysRevLett.129.047601} {\bibfield  {journal}
		{\bibinfo  {journal} {Physical Review Letters}\ }\textbf {\bibinfo {volume}
			{129}},\ \bibinfo {pages} {047601} (\bibinfo {year} {2022})}\BibitemShut
	{NoStop}%
	\bibitem [{\citenamefont {Kumar}\ \emph {et~al.}(2022)\citenamefont {Kumar},
		\citenamefont {Hu}, \citenamefont {MacDonald},\ and\ \citenamefont
		{Potter}}]{kumar_gate-tunable_2022}%
	\BibitemOpen
	\bibfield  {author} {\bibinfo {author} {\bibfnamefont {A.}~\bibnamefont
			{Kumar}}, \bibinfo {author} {\bibfnamefont {N.~C.}\ \bibnamefont {Hu}},
		\bibinfo {author} {\bibfnamefont {A.~H.}\ \bibnamefont {MacDonald}},\ and\
		\bibinfo {author} {\bibfnamefont {A.~C.}\ \bibnamefont {Potter}},\ }\bibfield
	{title} {{\bibinfo {title} {Gate-tunable heavy fermion
				quantum criticality in a moiré {Kondo} lattice}},\ }\href
	{https://doi.org/10.1103/PhysRevB.106.L041116} {\bibfield  {journal}
		{\bibinfo  {journal} {Physical Review B}\ }\textbf {\bibinfo {volume}
			{106}},\ \bibinfo {pages} {L041116} (\bibinfo {year} {2022})}\BibitemShut
	{NoStop}%
	\bibitem [{\citenamefont {Ramires}\ and\ \citenamefont
		{Lado}(2018)}]{ramires_electrically_2018}%
	\BibitemOpen
	\bibfield  {author} {\bibinfo {author} {\bibfnamefont {A.}~\bibnamefont
			{Ramires}}\ and\ \bibinfo {author} {\bibfnamefont {J.~L.}\ \bibnamefont
			{Lado}},\ }\bibfield  {title} {{\bibinfo {title}
			{Electrically {Tunable} {Gauge} {Fields} in {Tiny}-{Angle} {Twisted}
				{Bilayer} {Graphene}}},\ }\href
	{https://doi.org/10.1103/PhysRevLett.121.146801} {\bibfield  {journal}
		{\bibinfo  {journal} {Physical Review Letters}\ }\textbf {\bibinfo {volume}
			{121}},\ \bibinfo {pages} {146801} (\bibinfo {year} {2018})}\BibitemShut
	{NoStop}%
	\bibitem [{\citenamefont {Coleman}(1983)}]{coleman_1_1983}%
	\BibitemOpen
	\bibfield  {author} {\bibinfo {author} {\bibfnamefont {P.}~\bibnamefont
			{Coleman}},\ }\bibfield  {title} {{\bibinfo {title} {1
				{N} expansion for the {Kondo} lattice}},\ }\href
	{https://doi.org/10.1103/PhysRevB.28.5255} {\bibfield  {journal} {\bibinfo
			{journal} {Physical Review B}\ }\textbf {\bibinfo {volume} {28}},\ \bibinfo
		{pages} {5255} (\bibinfo {year} {1983})}\BibitemShut {NoStop}%
	\bibitem [{\citenamefont {Read}\ and\ \citenamefont
		{Newns}(1983)}]{read_solution_1983}%
	\BibitemOpen
	\bibfield  {author} {\bibinfo {author} {\bibfnamefont {N.}~\bibnamefont
			{Read}}\ and\ \bibinfo {author} {\bibfnamefont {D.~M.}\ \bibnamefont
			{Newns}},\ }\bibfield  {title} {\bibinfo {title} {On the solution of the
			{Coqblin}-{Schreiffer} {Hamiltonian} by the large-{N} expansion technique},\
	}\href {https://doi.org/10.1088/0022-3719/16/17/014} {\bibfield  {journal}
		{\bibinfo  {journal} {Journal of Physics C: Solid State Physics}\ }\textbf
		{\bibinfo {volume} {16}},\ \bibinfo {pages} {3273} (\bibinfo {year}
		{1983})}\BibitemShut {NoStop}%
	\bibitem [{\citenamefont {Kogut}(1979)}]{kogut_introduction_1979}%
	\BibitemOpen
	\bibfield  {author} {\bibinfo {author} {\bibfnamefont {J.~B.}\ \bibnamefont
			{Kogut}},\ }\bibfield  {title} {{\bibinfo {title} {An
				introduction to lattice gauge theory and spin systems}},\ }\href
	{https://doi.org/10.1103/RevModPhys.51.659} {\bibfield  {journal} {\bibinfo
			{journal} {Reviews of Modern Physics}\ }\textbf {\bibinfo {volume} {51}},\
		\bibinfo {pages} {659} (\bibinfo {year} {1979})}\BibitemShut {NoStop}%
	\bibitem [{\citenamefont {Senthil}\ \emph {et~al.}(2003)\citenamefont
		{Senthil}, \citenamefont {Vojta},\ and\ \citenamefont {Sachdev}}]{vojta}%
	\BibitemOpen
	\bibfield  {author} {\bibinfo {author} {\bibfnamefont {T.}~\bibnamefont
			{Senthil}}, \bibinfo {author} {\bibfnamefont {M.}~\bibnamefont {Vojta}},\
		and\ \bibinfo {author} {\bibfnamefont {S.}~\bibnamefont {Sachdev}},\
	}\bibfield  {title} {\bibinfo {title} {{\sl Fractionalized Fermi Liquids}},\
	}\href@noop {} {\bibfield  {journal} {\bibinfo  {journal} {Phys. Rev. Lett}\
		}\textbf {\bibinfo {volume} {90}},\ \bibinfo {pages} {216403} (\bibinfo
		{year} {2003})}\BibitemShut {NoStop}%
	\bibitem [{\citenamefont {Coleman}\ \emph {et~al.}(2005)\citenamefont
		{Coleman}, \citenamefont {Marston},\ and\ \citenamefont
		{Schofield}}]{Coleman:849264}%
	\BibitemOpen
	\bibfield  {author} {\bibinfo {author} {\bibfnamefont {P.}~\bibnamefont
			{Coleman}}, \bibinfo {author} {\bibfnamefont {J.~B.}\ \bibnamefont
			{Marston}},\ and\ \bibinfo {author} {\bibfnamefont {A.~J.}\ \bibnamefont
			{Schofield}},\ }\bibfield  {title} {\bibinfo {title} {{\sl Transport
				anomalies in a simplified model for a heavy electron quantum critical
				point}},\ }\href@noop {} {\bibfield  {journal} {\bibinfo  {journal} {Phys.
				Rev. B}\ }\textbf {\bibinfo {volume} {72}},\ \bibinfo {pages} {245111. 15 p}
		(\bibinfo {year} {2005})}\BibitemShut {NoStop}%
	\bibitem [{\citenamefont {Kitaev}(2006)}]{kitaev_anyons_2006}%
	\BibitemOpen
	\bibfield  {author} {\bibinfo {author} {\bibfnamefont {A.}~\bibnamefont
			{Kitaev}},\ }\bibfield  {title} {{\bibinfo {title}
			{Anyons in an exactly solved model and beyond}},\ }\href
	{https://doi.org/10.1016/j.aop.2005.10.005} {\bibfield  {journal} {\bibinfo
			{journal} {Annals of Physics}\ }\textbf {\bibinfo {volume} {321}},\ \bibinfo
		{pages} {2} (\bibinfo {year} {2006})}\BibitemShut {NoStop}%
	\bibitem [{\citenamefont {Yao}\ and\ \citenamefont
		{Lee}(2011)}]{yao_fermionic_2011}%
	\BibitemOpen
	\bibfield  {author} {\bibinfo {author} {\bibfnamefont {H.}~\bibnamefont
			{Yao}}\ and\ \bibinfo {author} {\bibfnamefont {D.-H.}\ \bibnamefont {Lee}},\
	}\bibfield  {title} {{\bibinfo {title} {Fermionic
				{Magnons}, {Non}-{Abelian} {Spinons}, and the {Spin} {Quantum} {Hall}
				{Effect} from an {Exactly} {Solvable} {Spin}- 1 / 2 {Kitaev} {Model} with
				{SU}(2) {Symmetry}}},\ }\href
	{https://doi.org/10.1103/PhysRevLett.107.087205} {\bibfield  {journal}
		{\bibinfo  {journal} {Physical Review Letters}\ }\textbf {\bibinfo {volume}
			{107}},\ \bibinfo {pages} {087205} (\bibinfo {year} {2011})}\BibitemShut
	{NoStop}%
	\bibitem [{\citenamefont {Hermanns}\ and\ \citenamefont
		{Trebst}(2014)}]{Hermanns}%
	\BibitemOpen
	\bibfield  {author} {\bibinfo {author} {\bibfnamefont {M.}~\bibnamefont
			{Hermanns}}\ and\ \bibinfo {author} {\bibfnamefont {S.}~\bibnamefont
			{Trebst}},\ }\bibfield  {title} {\bibinfo {title} {Quantum spin liquid with a
			majorana fermi surface on the three-dimensional hyperoctagon lattice},\
	}\href {https://doi.org/10.1103/PhysRevB.89.235102} {\bibfield  {journal}
		{\bibinfo  {journal} {Phys. Rev. B}\ }\textbf {\bibinfo {volume} {89}},\
		\bibinfo {pages} {235102} (\bibinfo {year} {2014})}\BibitemShut {NoStop}%
	\bibitem [{\citenamefont {Hermanns}\ \emph {et~al.}(2018)\citenamefont
		{Hermanns}, \citenamefont {Kimchi},\ and\ \citenamefont
		{Knolle}}]{hermanns_physics_2018}%
	\BibitemOpen
	\bibfield  {author} {\bibinfo {author} {\bibfnamefont {M.}~\bibnamefont
			{Hermanns}}, \bibinfo {author} {\bibfnamefont {I.}~\bibnamefont {Kimchi}},\
		and\ \bibinfo {author} {\bibfnamefont {J.}~\bibnamefont {Knolle}},\
	}\bibfield  {title} {{\bibinfo {title} {Physics of the
				{Kitaev} {Model}: {Fractionalization}, {Dynamic} {Correlations}, and
				{Material} {Connections}}},\ }\href
	{https://doi.org/10.1146/annurev-conmatphys-033117-053934} {\bibfield
		{journal} {\bibinfo  {journal} {Annual Review of Condensed Matter Physics}\
		}\textbf {\bibinfo {volume} {9}},\ \bibinfo {pages} {17} (\bibinfo {year}
		{2018})}\BibitemShut {NoStop}%
	\bibitem [{\citenamefont {O'Brien}\ \emph {et~al.}(2016)\citenamefont
		{O'Brien}, \citenamefont {Hermanns},\ and\ \citenamefont
		{Trebst}}]{obrien_classification_2016}%
	\BibitemOpen
	\bibfield  {author} {\bibinfo {author} {\bibfnamefont {K.}~\bibnamefont
			{O'Brien}}, \bibinfo {author} {\bibfnamefont {M.}~\bibnamefont {Hermanns}},\
		and\ \bibinfo {author} {\bibfnamefont {S.}~\bibnamefont {Trebst}},\
	}\bibfield  {title} {{\bibinfo {title} {Classification of
				gapless {Z} 2 spin liquids in three-dimensional {Kitaev} models}},\ }\href
	{https://doi.org/10.1103/PhysRevB.93.085101} {\bibfield  {journal} {\bibinfo
			{journal} {Physical Review B}\ }\textbf {\bibinfo {volume} {93}},\ \bibinfo
		{pages} {085101} (\bibinfo {year} {2016})}\BibitemShut {NoStop}%
	\bibitem [{\citenamefont {Komijani}\ \emph {et~al.}(2018)\citenamefont
		{Komijani}, \citenamefont {Toth}, \citenamefont {Chandra},\ and\
		\citenamefont {Coleman}}]{komijani_order_2018}%
	\BibitemOpen
	\bibfield  {author} {\bibinfo {author} {\bibfnamefont {Y.}~\bibnamefont
			{Komijani}}, \bibinfo {author} {\bibfnamefont {A.}~\bibnamefont {Toth}},
		\bibinfo {author} {\bibfnamefont {P.}~\bibnamefont {Chandra}},\ and\ \bibinfo
		{author} {\bibfnamefont {P.}~\bibnamefont {Coleman}},\ }\bibfield  {title}
	{\bibinfo {title} {Order {Fractionalization}}\ }\href
	{https://doi.org/10.48550/ARXIV.1811.11115} {10.48550/ARXIV.1811.11115}
	(\bibinfo {year} {2018}),\ \bibinfo {note} {publisher: arXiv Version Number:
		2}\BibitemShut {NoStop}%
	\bibitem [{\citenamefont {Seifert}\ \emph {et~al.}(2020)\citenamefont
		{Seifert}, \citenamefont {Dong}, \citenamefont {Chulliparambil},
		\citenamefont {Vojta}, \citenamefont {Tu},\ and\ \citenamefont
		{Janssen}}]{janssen20}%
	\BibitemOpen
	\bibfield  {author} {\bibinfo {author} {\bibfnamefont {U.~F.~P.}\
			\bibnamefont {Seifert}}, \bibinfo {author} {\bibfnamefont {X.-Y.}\
			\bibnamefont {Dong}}, \bibinfo {author} {\bibfnamefont {S.}~\bibnamefont
			{Chulliparambil}}, \bibinfo {author} {\bibfnamefont {M.}~\bibnamefont
			{Vojta}}, \bibinfo {author} {\bibfnamefont {H.-H.}\ \bibnamefont {Tu}},\ and\
		\bibinfo {author} {\bibfnamefont {L.}~\bibnamefont {Janssen}},\ }\bibfield
	{title} {\bibinfo {title} {Fractionalized fermionic quantum criticality in
			spin-orbital mott insulators},\ }\href
	{https://doi.org/10.1103/PhysRevLett.125.257202} {\bibfield  {journal}
		{\bibinfo  {journal} {Phys. Rev. Lett.}\ }\textbf {\bibinfo {volume} {125}},\
		\bibinfo {pages} {257202} (\bibinfo {year} {2020})}\BibitemShut {NoStop}%
	\bibitem [{\citenamefont {Mishchenko}\ \emph {et~al.}(2017)\citenamefont
		{Mishchenko}, \citenamefont {Kato},\ and\ \citenamefont
		{Motome}}]{mishenko17}%
	\BibitemOpen
	\bibfield  {author} {\bibinfo {author} {\bibfnamefont {P.~A.}\ \bibnamefont
			{Mishchenko}}, \bibinfo {author} {\bibfnamefont {Y.}~\bibnamefont {Kato}},\
		and\ \bibinfo {author} {\bibfnamefont {Y.}~\bibnamefont {Motome}},\
	}\bibfield  {title} {\bibinfo {title} {Finite-temperature phase transition to
			a kitaev spin liquid phase on a hyperoctagon lattice: A large-scale quantum
			monte carlo study},\ }\href {https://doi.org/10.1103/PhysRevB.96.125124}
	{\bibfield  {journal} {\bibinfo  {journal} {Phys. Rev. B}\ }\textbf {\bibinfo
			{volume} {96}},\ \bibinfo {pages} {125124} (\bibinfo {year}
		{2017})}\BibitemShut {NoStop}%
	\bibitem [{\citenamefont {Eschmann}\ \emph {et~al.}(2020)\citenamefont
		{Eschmann}, \citenamefont {Mishchenko}, \citenamefont {O'Brien},
		\citenamefont {Bojesen}, \citenamefont {Kato}, \citenamefont {Hermanns},
		\citenamefont {Motome},\ and\ \citenamefont
		{Trebst}}]{eschmann_thermodynamic_2020}%
	\BibitemOpen
	\bibfield  {author} {\bibinfo {author} {\bibfnamefont {T.}~\bibnamefont
			{Eschmann}}, \bibinfo {author} {\bibfnamefont {P.~A.}\ \bibnamefont
			{Mishchenko}}, \bibinfo {author} {\bibfnamefont {K.}~\bibnamefont {O'Brien}},
		\bibinfo {author} {\bibfnamefont {T.~A.}\ \bibnamefont {Bojesen}}, \bibinfo
		{author} {\bibfnamefont {Y.}~\bibnamefont {Kato}}, \bibinfo {author}
		{\bibfnamefont {M.}~\bibnamefont {Hermanns}}, \bibinfo {author}
		{\bibfnamefont {Y.}~\bibnamefont {Motome}},\ and\ \bibinfo {author}
		{\bibfnamefont {S.}~\bibnamefont {Trebst}},\ }\bibfield  {title}
	{{\bibinfo {title} {Thermodynamic classification of
				three-dimensional {Kitaev} spin liquids}},\ }\href
	{https://doi.org/10.1103/PhysRevB.102.075125} {\bibfield  {journal} {\bibinfo
			{journal} {Physical Review B}\ }\textbf {\bibinfo {volume} {102}},\ \bibinfo
		{pages} {075125} (\bibinfo {year} {2020})}\BibitemShut {NoStop}%
	\bibitem [{\citenamefont {Hermanns}\ \emph {et~al.}(2015)\citenamefont
		{Hermanns}, \citenamefont {Trebst},\ and\ \citenamefont {Rosch}}]{Hermanns2}%
	\BibitemOpen
	\bibfield  {author} {\bibinfo {author} {\bibfnamefont {M.}~\bibnamefont
			{Hermanns}}, \bibinfo {author} {\bibfnamefont {S.}~\bibnamefont {Trebst}},\
		and\ \bibinfo {author} {\bibfnamefont {A.}~\bibnamefont {Rosch}},\ }\bibfield
	{title} {\bibinfo {title} {Spin-peierls instability of three-dimensional
			spin liquids with majorana fermi surfaces},\ }\href
	{https://doi.org/10.1103/PhysRevLett.115.177205} {\bibfield  {journal}
		{\bibinfo  {journal} {Phys. Rev. Lett.}\ }\textbf {\bibinfo {volume} {115}},\
		\bibinfo {pages} {177205} (\bibinfo {year} {2015})}\BibitemShut {NoStop}%
	\bibitem [{\citenamefont {Coleman}\ \emph {et~al.}(2001)\citenamefont
		{Coleman}, \citenamefont {Pépin}, \citenamefont {Si},\ and\ \citenamefont
		{Ramazashvili}}]{coleman_how_2001}%
	\BibitemOpen
	\bibfield  {author} {\bibinfo {author} {\bibfnamefont {P.}~\bibnamefont
			{Coleman}}, \bibinfo {author} {\bibfnamefont {C.}~\bibnamefont {Pépin}},
		\bibinfo {author} {\bibfnamefont {Q.}~\bibnamefont {Si}},\ and\ \bibinfo
		{author} {\bibfnamefont {R.}~\bibnamefont {Ramazashvili}},\ }\bibfield
	{title} {\bibinfo {title} {How do {Fermi} liquids get heavy and die?},\
	}\href {https://doi.org/10.1088/0953-8984/13/35/202} {\bibfield  {journal}
		{\bibinfo  {journal} {Journal of Physics: Condensed Matter}\ }\textbf
		{\bibinfo {volume} {13}},\ \bibinfo {pages} {R723} (\bibinfo {year}
		{2001})}\BibitemShut {NoStop}%
	\bibitem [{\citenamefont {Senthil}\ \emph {et~al.}(2004)\citenamefont
		{Senthil}, \citenamefont {Vojta},\ and\ \citenamefont
		{Sachdev}}]{senthil_weak_2004}%
	\BibitemOpen
	\bibfield  {author} {\bibinfo {author} {\bibfnamefont {T.}~\bibnamefont
			{Senthil}}, \bibinfo {author} {\bibfnamefont {M.}~\bibnamefont {Vojta}},\
		and\ \bibinfo {author} {\bibfnamefont {S.}~\bibnamefont {Sachdev}},\
	}\bibfield  {title} {{\bibinfo {title} {Weak magnetism
				and non-{Fermi} liquids near heavy-fermion critical points}},\ }\href
	{https://doi.org/10.1103/PhysRevB.69.035111} {\bibfield  {journal} {\bibinfo
			{journal} {Physical Review B}\ }\textbf {\bibinfo {volume} {69}},\ \bibinfo
		{pages} {035111} (\bibinfo {year} {2004})}\BibitemShut {NoStop}%
	\bibitem [{\citenamefont {Paschen}\ \emph {et~al.}(2004)\citenamefont
		{Paschen}, \citenamefont {Lühmann}, \citenamefont {Wirth}, \citenamefont
		{Gegenwart}, \citenamefont {Trovarelli}, \citenamefont {Geibel},
		\citenamefont {Steglich}, \citenamefont {Coleman},\ and\ \citenamefont
		{Si}}]{paschen_hall-effect_2004}%
	\BibitemOpen
	\bibfield  {author} {\bibinfo {author} {\bibfnamefont {S.}~\bibnamefont
			{Paschen}}, \bibinfo {author} {\bibfnamefont {T.}~\bibnamefont {Lühmann}},
		\bibinfo {author} {\bibfnamefont {S.}~\bibnamefont {Wirth}}, \bibinfo
		{author} {\bibfnamefont {P.}~\bibnamefont {Gegenwart}}, \bibinfo {author}
		{\bibfnamefont {O.}~\bibnamefont {Trovarelli}}, \bibinfo {author}
		{\bibfnamefont {C.}~\bibnamefont {Geibel}}, \bibinfo {author} {\bibfnamefont
			{F.}~\bibnamefont {Steglich}}, \bibinfo {author} {\bibfnamefont
			{P.}~\bibnamefont {Coleman}},\ and\ \bibinfo {author} {\bibfnamefont
			{Q.}~\bibnamefont {Si}},\ }\bibfield  {title} {{\bibinfo
			{title} {Hall-effect evolution across a heavy-fermion quantum critical
				point}},\ }\href {https://doi.org/10.1038/nature03129} {\bibfield  {journal}
		{\bibinfo  {journal} {Nature}\ }\textbf {\bibinfo {volume} {432}},\ \bibinfo
		{pages} {881} (\bibinfo {year} {2004})}\BibitemShut {NoStop}%
	\bibitem [{\citenamefont {Lammert}\ \emph {et~al.}(1993)\citenamefont
		{Lammert}, \citenamefont {Rokhsar},\ and\ \citenamefont
		{Toner}}]{lammert_topology_1993}%
	\BibitemOpen
	\bibfield  {author} {\bibinfo {author} {\bibfnamefont {P.~E.}\ \bibnamefont
			{Lammert}}, \bibinfo {author} {\bibfnamefont {D.~S.}\ \bibnamefont
			{Rokhsar}},\ and\ \bibinfo {author} {\bibfnamefont {J.}~\bibnamefont
			{Toner}},\ }\bibfield  {title} {{\bibinfo {title}
			{Topology and nematic ordering}},\ }\href
	{https://doi.org/10.1103/PhysRevLett.70.1650} {\bibfield  {journal} {\bibinfo
			{journal} {Physical Review Letters}\ }\textbf {\bibinfo {volume} {70}},\
		\bibinfo {pages} {1650} (\bibinfo {year} {1993})}\BibitemShut {NoStop}%
	\bibitem [{\citenamefont {Lammert}\ \emph {et~al.}(1995)\citenamefont
		{Lammert}, \citenamefont {Rokhsar},\ and\ \citenamefont
		{Toner}}]{lammert_topology_1995}%
	\BibitemOpen
	\bibfield  {author} {\bibinfo {author} {\bibfnamefont {P.~E.}\ \bibnamefont
			{Lammert}}, \bibinfo {author} {\bibfnamefont {D.~S.}\ \bibnamefont
			{Rokhsar}},\ and\ \bibinfo {author} {\bibfnamefont {J.}~\bibnamefont
			{Toner}},\ }\bibfield  {title} {{\bibinfo {title}
			{Topology and nematic ordering. {I}. {A} gauge theory}},\ }\href
	{https://doi.org/10.1103/PhysRevE.52.1778} {\bibfield  {journal} {\bibinfo
			{journal} {Physical Review E}\ }\textbf {\bibinfo {volume} {52}},\ \bibinfo
		{pages} {1778} (\bibinfo {year} {1995})}\BibitemShut {NoStop}%
	\bibitem [{\citenamefont {Sachdev}(2019)}]{sachdev_topological_2019}%
	\BibitemOpen
	\bibfield  {author} {\bibinfo {author} {\bibfnamefont {S.}~\bibnamefont
			{Sachdev}},\ }\bibfield  {title} {\bibinfo {title} {Topological order,
			emergent gauge fields, and {Fermi} surface reconstruction},\ }\href
	{https://doi.org/10.1088/1361-6633/aae110} {\bibfield  {journal} {\bibinfo
			{journal} {Reports on Progress in Physics}\ }\textbf {\bibinfo {volume}
			{82}},\ \bibinfo {pages} {014001} (\bibinfo {year} {2019})}\BibitemShut
	{NoStop}%
	\bibitem [{\citenamefont {Fradkin}\ and\ \citenamefont
		{Shenker}(1979)}]{fradkin_phase_1979}%
	\BibitemOpen
	\bibfield  {author} {\bibinfo {author} {\bibfnamefont {E.}~\bibnamefont
			{Fradkin}}\ and\ \bibinfo {author} {\bibfnamefont {S.~H.}\ \bibnamefont
			{Shenker}},\ }\bibfield  {title} {{\bibinfo {title}
			{Phase diagrams of lattice gauge theories with {Higgs} fields}},\ }\href
	{https://doi.org/10.1103/PhysRevD.19.3682} {\bibfield  {journal} {\bibinfo
			{journal} {Physical Review D}\ }\textbf {\bibinfo {volume} {19}},\ \bibinfo
		{pages} {3682} (\bibinfo {year} {1979})}\BibitemShut {NoStop}%
	\bibitem [{\citenamefont {Panigrahi}\ \emph {et~al.}(2023)\citenamefont
		{Panigrahi}, \citenamefont {Coleman},\ and\ \citenamefont
		{Tsvelik}}]{panigrahi_analytic_2023}%
	\BibitemOpen
	\bibfield  {author} {\bibinfo {author} {\bibfnamefont {A.}~\bibnamefont
			{Panigrahi}}, \bibinfo {author} {\bibfnamefont {P.}~\bibnamefont {Coleman}},\
		and\ \bibinfo {author} {\bibfnamefont {A.}~\bibnamefont {Tsvelik}},\
	}\bibfield  {title} {{\bibinfo {title} {Analytic
				calculation of the vison gap in the {Kitaev} spin liquid}},\ }\href
	{https://doi.org/10.1103/PhysRevB.108.045151} {\bibfield  {journal} {\bibinfo
			{journal} {Physical Review B}\ }\textbf {\bibinfo {volume} {108}},\ \bibinfo
		{pages} {045151} (\bibinfo {year} {2023})}\BibitemShut {NoStop}%
	\bibitem [{\citenamefont {Fradkin}(2021)}]{fradkin_quantum_2021}%
	\BibitemOpen
	\bibfield  {author} {\bibinfo {author} {\bibfnamefont {E.}~\bibnamefont
			{Fradkin}},\ }\href@noop {} {\emph {\bibinfo {title} {Quantum field theory:
				an integrated approach}}}\ (\bibinfo  {publisher} {Princeton University
		Press},\ \bibinfo {address} {Princeton},\ \bibinfo {year} {2021})\BibitemShut
	{NoStop}%
\end{thebibliography}
\end{document}